\documentclass[useAMS,usenatbib]{mn2e}
\usepackage{tikz}
\usepackage{amssymb}
\usepackage[normalem]{ulem}
\usepackage{cancel}
\usepackage{amsmath}

\newcommand{\vc}{v_{\rm c}}  
\newcommand{\mt}{m_{\rm t}}  
\newcommand{\mpr}{m_{\rm p}}  
\newcommand{\Ediss}{E_{\rm diss}}  
\newcommand{\Einit}{E_{\rm init}}  
\newcommand{\Rp}{R_{\rm p}}  
\newcommand{\Rt}{R_{\rm t}}  
\newcommand{\rhomatrix}{\rho_{\rm matrix}}  
\newcommand{\SiOtwo}{{\rm SiO_2}}  
\newcommand{\Edissmax}{E_{\rm diss,max}}  
\usepackage{cancel}

\newcommand{\pmean}{p_{\rm m}}				

\newcommand{\phirbd}{\phi_{\rm RBD}}			

\newcommand{\mproj}{m_{\rm p}}

\title[Fragmentation velocities for cm-sized dust collisions]{Growth and fragmentation of centimetre-sized dust aggregates: the dependence on aggregate size and porosity}
\author[Meru et. al]{\parbox{\textwidth}{Farzana Meru$^{1,2}$\thanks{farzana.meru@phys.ethz.ch}, Ralf J. Geretshauser$^2$, Christoph Sch\"afer$^2$, Roland Speith$^3$ and Wilhelm Kley$^2$}\vspace{0.5cm}\\
$^1$Institut f\"ur Astronomie, ETH Z\"urich, Wolfgang-Pauli-Strasse 27, 8093 Z\"urich, Switzerland\\
$^2$Institut f\"ur Astronomie und Astrophysik, Universit\"at T\"ubingen, Auf der Morgenstelle 10, 72076 T\"ubingen, Germany\\
$^3$Physikalisches Institut, Universit\"at T\"ubingen, Auf der Morgenstelle 14, 72076 T\"ubingen, Germany}
\begin{document}



\maketitle

\label{firstpage}

\begin{abstract}
We carry out three-dimensional Smoothed Particle Hydrodynamics simulations of spherical homogeneous $\SiOtwo$ dust aggregates to investigate how the mass and the porosity of the aggregates affects their ability to survive an impact at various different collision velocities (between $1-27.5$~m/s).  We explore how the threshold velocities for fragmentation vary with these parameters.  Crucially, we find that the porosity plays a part of utmost importance in determining the outcome of collisions.  In particular, we find that aggregates with filling factors $\gtrsim 37\%$ are significantly weakened and that the velocity regime in which the aggregates grow is reduced or even non-existent (instead, the aggregates either rebound off each other or break apart).  At filling factors less than $\approx 37\%$ we find that more porous objects are weaker but not as weak as highly compact objects with filling factors $\gtrsim 37\%$.  In addition we find that (for a given aggregate density) collisions between very different mass objects have higher threshold velocities than those between very similar mass objects.  We find that fragmentation velocities are higher than the typical values of $1$~m/s and that growth can even occur for velocities as high as 27.5~m/s.  Therefore, while the growth of aggregates is more likely if collisions between different sized objects occurs or if the aggregates are porous with filling factor $< 37\%$, it may also be hindered if the aggregates become too compact.
\end{abstract}


\section{Introduction}
\label{sec:intro}
A key question in planet formation is how objects may grow from small sizes (micron to centimetre sized dust aggregates) to planetary sized objects.  Observations are able to shed light on the small sizes (up to $\approx$~cm sizes; \citealp[e.g.][]{Wilner2005_cm,Rodman_etal_mm_grains,Lommen_mm_cm,Ricci_mm_cm_rhoOph}; also see Section 6.3.3 of \citealp{Williams_Cieza_disc_review}) while the vast number of observations of extra-solar planets provide us with the properties of the latter (http://exoplanet.eu).  However, we are blind to the size regime in between, i.e. to approximately eight orders of magnitude.  Understanding the planet formation process is therefore challenging as clues are pieced together to infer the growth process.

To aid the understanding, a number of coagulation and fragmentation simulations have been carried out, involving both the gas and dust dynamics, to explore the growth of dust aggregates from micron to larger sizes \citep[e.g.][]{Ormel_porosity,Brauer_coag_frag,Suyama_porosity_model,Okuzumi_porosity_model,Zsom_bouncing,Birnstiel_coag_frag,Windmark_lucky_ptcl,Okuzumi_vel_pdf,Windmark_vel_pdf,Garaud_vel_pdf,BD_discs_letter}.  However, such codes require an understanding of the collisional outcome of aggregates in order to model their coagulation accurately.  In addition, not only is it important to know when aggregates will collide and grow, but it is also useful to know when the collisional outcome is an obstacle to their growth.  In particular, if collision velocities are too high, aggregates will certainly break apart and fragment.  But knowing when this will happen is not so well explored.

Earlier experimental work by \cite{Blum_Munch_1m_s} using silicates ($\rm ZrSiO_4$ and $\SiOtwo$) suggested that the threshold velocity for fragmentation was $\approx 1-4$~m/s.  Though they do indicate that composition and porosity may play a part, a threshold velocity of 1m/s is often taken as the standard result.  In a protoplanetary disc, particle velocities may vary from below 1m/s to $\approx 100$~m/s for $\approx 1-10$~cm sized objects \citep{Weidenschilling1977} though this is also dependent on the disc parameters adopted.  If a threshold velocity of 1~m/s is always used, this would imply that it is very difficult for aggregates to grow in a disc.  Other parameters that may be important include the impact parameter (\citealp{Wada_icy_collisions}; \citealp{Ringl_vth}), rotation, aggregate mass and inhomogeneity \citep{Geretshauser_inhomogeneity}.  Thus it would be surprising if a single threshold velocity for fragmentation existed.

\cite{Wurm_25m/s_impacts} carried out laboratory experiments of mm-sized projectiles colliding with cm-sized targets, both using $\SiOtwo$ dust with porosities of $\approx 66\%$.  They found that growth can occur at collision velocities as high as 25m/s.  More recently, \cite{Teiser_Wurm_highVcoll} carried out laboratory experiments and showed that for very unequal sized aggregates ($\approx 0.5$~mm sized projectile and a target that was at least ten times larger) growth can occur following a collision even when the collional velocity is as high as $\approx 55$~m/s.  Most coagulation-fragmentation codes have not taken the aggregate masses into account when considering the velocity threshold for fragmentation and have taken the threshold value of 1m/s as a hard limit beyond which fragmentation occurs, though recent attempts by \cite{Windmark_lucky_ptcl} have started to consider the projectile mass and to a certain extent the target mass.

\cite{Ormel_porosity}, \cite{Suyama_porosity_model} and \cite{Okuzumi_porosity_model} developed porosity models to understand the growth of aggregates.  It is clear from all these studies that the effect of changing the porosity affects the dynamics of the aggregates and thus their growth.  In addition, \cite{Okuzumi_RD_growth_Stokes} used the model by \cite{Okuzumi_porosity_model} to show that growth can occur rapidly to overcome the radial drift barrier.  \cite{Zsom_bouncing} implemented a simple procedure to model the porosity evolution (using the experimental results by \citealp{Guttler.2010}) to determine the growth of dust locally in a disc.  However, since these experiments covered a small parameter space (in comparison to the large parameter space needed to be explored) a more detailed understanding of the effects that the porosity evolution would have on the fragmentation of particles is needed.  Therefore, understanding how the threshold velocities for fragmentation are affected by the aggregate porosity will enable us to recognise the limitations to growth.  One would expect that the structure of an aggregate will play a big part in its ability to survive an impact.

In a collision, the energy will first go into the elastic loading of the aggregate.  Unless the collision is purely elastic, some energy will be dissipated.  The onset of any deformation is always elastic then becomes plastic.  For macroscopic (cm-sized) objects plastic deformation sets in at a much lower collision velocity than for smaller aggregates \citep{Chokshi_vs}.  Therefore, plastic deformation is more important to consider for larger objects than small-sized objects.  Aggregates combine together if they are able to dissipate as much energy as possible.  When aggregates grow, some of the energy is released via plastic deformation which then allows the aggregates to coalesce.  The remainder of the energy goes into the kinetic energy of the final aggregate.  If there is simply too much collisional energy to get rid of and if the aggregates cannot do this via plastic deformation, the energy is released through kinetic energy of smaller fragments, i.e. they break apart.  This process will depend on the structure of the aggregates and hence their porosities.

While laboratory experiments have looked at collisional outcomes in the past, the regime in which they tend to focus on is the size range up to centimetre sizes \citep[see Section 5 of][for a comprehensive review]{Blum_Wurm_review}.  Experiments are only just beginning to explore the decimetre range \citep{Teiser_Wurm_decimeter_growth} (though these typically involve decimetre-sized plates as targets and smaller projectiles).  Furthermore, a controlled experiment is much more demanding in the laboratory since the preparation of the samples may affect the properties of the aggregate such as their size and porosity.  Moreover, a full parameter study in the laboratory is not possible.  Therefore, a numerical parameter study is much easier, cheaper and faster.

We carry out three-dimensional Smoothed Particle Hydrodynamics simulations of solid body collisions of spherical porous $\SiOtwo$ dust aggregates to determine how the threshold velocity for fragmentation varies with the porosity and size of the aggregates.  The outcome of collisions can be characterised by using the Four-Population model presented by \cite{Four_population}, which involves the properties of the largest object, the second largest object, a power-law population and a sub-resolution population.  In the context of this model, we mostly focus on the first group, i.e. the final mass of the largest fragment, since we are concerned with how this compares to the initial target mass, i.e. whether the aggregate grows or not.

Section~\ref{sec:analytics} describes the analytical expectations for a perfect sticking scenario.  Sections~\ref{sec:numerics} and~\ref{sec:sim} describe the numerical method employed and the simulations carried out, respectively.  Section~\ref{sec:results} presents the results.  We then discuss and make conclusions in Sections~\ref{sec:disc} and~\ref{sec:conc}, respectively.

\section{Analytical expectations for perfect sticking}
\label{sec:analytics}

When a target of mass, $\mt$, and a projectile of mass $\mpr$, collide with a velocity, $\vc$, \emph{and} if the aggregates stick together then energy will be dissipated.  The maximum energy that can be dissipated, $\Ediss$, in any collision occurs when there is perfect sticking, such that all the initial centre of mass kinetic energy, $\Einit$, is dissipated.  By conservation of energy and by considering the motion in the centre of mass frame, the dissipated energy in a perfect sticking collision is

\begin{equation}
\Edissmax = \Einit = \frac{1}{2}\frac{\mt \mpr \vc^2}{(\mt + \mpr)} = \frac{1}{2}
\mpr \vc^2 \left ( 1 + \frac{\mpr}{\mt} \right )^{-1}.
\label{eq:Ediss}
\end{equation}

The filling factor of a dust aggregate is defined as

\begin{equation}
\phi = \frac{\rho}{\rhomatrix} = 1 - \Phi,
\end{equation}
where $\Phi$ and $\rho$ are the porosity and density of the aggregate and $\rho_{\rm matrix}$ is the density of the matrix material.  Equation~\ref{eq:Ediss} can be written in terms of the aggregate filling factor as follows (for a projectile and target of the same filling factor):

\begin{align}
\Edissmax & = \frac{2}{3} \pi \phi \rho_{\rm matrix} \vc^2 \frac{\Rp^3 \Rt^3}{(\Rp^3+\Rt^3)} \nonumber \\
& = \frac{2}{3} \pi \Rp^3 \phi \rho_{\rm matrix} \vc^2 \left [ 1 + \left ( \frac{\Rp}{\Rt} \right )^3 \right ]^{-1}
\label{eq:Ediss_phi}
\end{align}
where $\Rt$ and $\Rp$ are the target and projectile radii, respectively.  Therefore as the filling factor of the aggregate increases, the aggregate's ability to dissipate energy also increases due to its increased mass.

\section{Numerical method}
\label{sec:numerics} 

\begin{figure}
\centering
  \includegraphics[width=1.0\columnwidth]{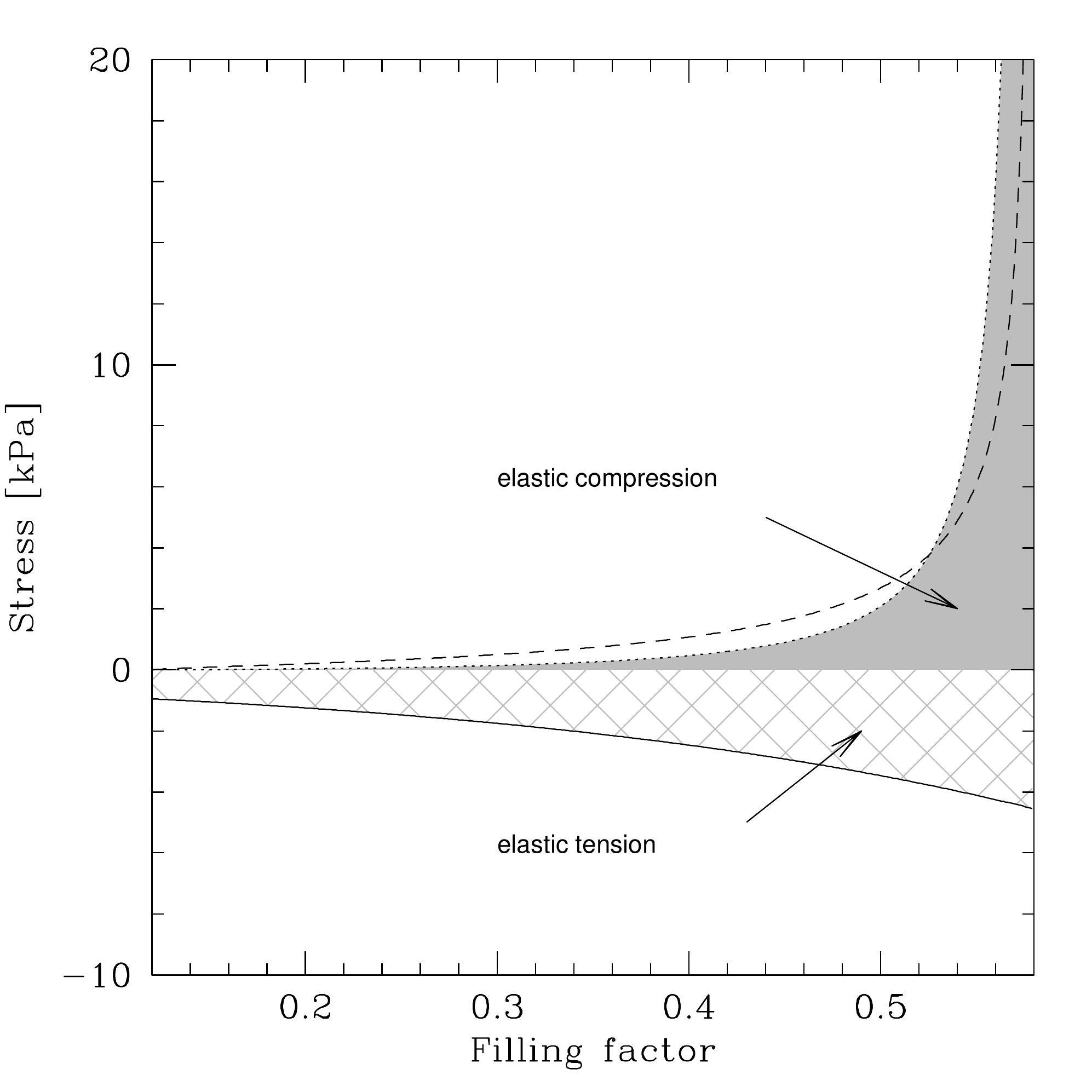}
  \caption{Graph showing how the tensile (solid line), compressive (dotted line) and shear (dashed line) strengths change with filling factor.  The strengths increase in magnitude as the filling factor increases.  The grey shaded and hashed regions show where elastic compression and tension, respectively, take place.}
\label{fig:strengths}
\end{figure}

We use a parallel Smoothed Particle Hydrodynamics (SPH) code developed by \cite{parasph} ({\sc parasph}), and expanded to simulate solid bodies \citep{Schaefer_SPH_ice} and porous material \citep{Geretshauser.2010}.  The code, including the porosity model, has been calibrated with carefully controlled laboratory experiments \citep{Guttler.2009,Geretshauser.2010} for $\SiOtwo$ dust aggregates (with $\rhomatrix = 2 \rm g/cm^3$) \citep{Blum_rhomatrix} and models both the elastic and plastic behaviour.  It is capable of modelling the collision of dust aggregates in the strength-dominated regime (i.e. small dust aggregates that are unaffected by their self-gravity).  The code takes into account the filling factor dependent shear, compressive and tensile strengths as well as the bulk modulus of the material.  The bulk modulus is given by

\begin{equation}
K (\phi) = K_0 \left( \frac{\phi}{\phirbd} \right)^\gamma,
\label{eq:bulk-modulus}
\end{equation}
where $K_0 = 4.5$~kPa, $\gamma = 4$ and $\phirbd =0.15$ is the filling factor of the uncompressed dust material.  The tensile and compressive strengths are given by

\begin{equation}
T(\phi) = - 10^{a + b\phi}  \rm Pa
\label{eq:tensile-strength}
\end{equation} 
and

\begin{equation}
\Sigma(\phi) = \pmean \left( \frac{\phi_2 - \phi_1}{\phi_2 - \phi} - 1 \right)^{\Delta\,\ln{10}},
\label{eq:compressive-strength}
\end{equation}
respectively, where $\phi_1 = 0.12$ and $\phi_2 = 0.58$ define the porosity limits of $\SiOtwo$ dust aggregates, $\pmean = 0.26$~kPa, $\Delta = 0.58$, $a = 2.8$ and $b = 1.48$.  The values for $K_0$, $\gamma$, $a$, $b$, $\phi_{\rm min}$, $\phi_{\rm max}$, $\pmean$ and $\Delta$ are values that were calibrated by means of laboratory experiments \citep{Guttler.2009,Geretshauser.2010}.  The stress-strain relation is usually given by

\begin{equation}
\sigma_{\alpha \beta} = -p \delta_{\alpha \beta} + S_{\alpha \beta},
\end{equation}
where the first term represents the hydrostatic stress (taking into account the bulk modulus in the elastic regime and the compressive and tensile strengths in the plastic regime), and the second term considers the deviatoric stress tensor, $S_{\alpha \beta}$, which takes into account the shear strength, $Y(\phi)$, given by

\begin{equation}
Y(\phi) = \sqrt{\Sigma(\phi) | T(\phi) |}.
\label{eq:shear-strength}
\end{equation}
In the elastic regime the stress and strain are related to each other via Hooke's law.  However, in a perfectly plastic regime the stress stays constant when the strain is changed.  Therefore, to model plasticity a deviation from the stress-strain relation from Hooke's law is necessary and a reduction in the deviatoric stress is needed.  The \cite{Von_Mises} yield criterion states that plasticity occurs when the shear strength is less than the yield stress.  In mathematical terms, plasticity occurs when $Y^2/3 < J_2$, where $J_2$ is the second invariant of the deviatoric stress tensor given by $J_2 = \frac{1}{2} S_{\alpha \beta} S_{\alpha \beta}$ and is the square of the yield stress of the material in pure shear.  Following \cite{Benz_Asphaug_reduce_shear}, our implementation changes the stress according to

\begin{equation}
S_{\alpha \beta} \rightarrow f S_{\alpha \beta}
\end{equation}
where $f = {\rm min}[1, Y^2/ (3 J_2)]$ is a factor which takes into account both the elastic and plastic regimes.  In the elastic regime, $f = 1$.

The material behaves elastically when the shear strength in the material is greater than its yield stress (i.e. $3 J_2 < Y^2$) and when either of the following occur:

1) during compression when the pressure is less than the compressive strength ($p<\Sigma(\phi)$; see Figure~\ref{fig:strengths}), or

2) during tension when the magnitude of the pressure is smaller than the magnitude of the tensile strength ($|p| < | T(\phi) |$; see Figure~\ref{fig:strengths}).

In the elastic regime, the hydrostatic pressure is given by \citep{Sirono2004}

\begin{equation}
p (\phi) = K(\phi') \left ( \frac{\phi}{\phi'} - 1 \right ).
\end{equation}
where $\phi'$ is the reference filling factor i.e. the filling factor of the material under the influence of zero external stress.  For any given material with a filling factor under no external force, there exist critical filling factors above ($\phi_+$) and below ($\phi_-$) this, beyond which the material behaves plastically.  If $\phi > \phi_+$ then the material is under compression and the pressure is given by the compressive strength.  If $\phi < \phi_-$, the material is under tension and the pressure is given by the tensile strength.  In addition, the material also behaves plastically when the \cite{Von_Mises} yield criterion is satisfied, as described above.

The hydrostatic equation of state, $p$, can therefore be summarised as \citep[see][for details]{Geretshauser.2010}:

\begin{equation}
p(\phi) =
\left\{
\begin{array}{l l l}
    T(\phi) & \phi < \phi_- & {\rm plastic~regime}, \\[1.25ex]
    K(\phi') \left ( \frac{\phi}{\phi'} - 1 \right ) & \phi_- \le \phi \le \phi_+ & {\rm elastic~regime}, \\[1.25ex]
    \Sigma(\phi) &  \phi > \phi_+ & {\rm plastic~regime}.
\end{array}
\right.
\label{eq:hydro_eos}
\end{equation}
We stress that the critical filling factors strongly depend on the reference filling factor (which decreases in the tensile regime and increases in the compressive regime) so that the equation of state changes based on the reference filling factor in any one location of the dust aggregate.

The implementation of the equation of state adopted here is particularly suitable for modelling $\SiOtwo$ where the strengths increase as the filling factor increases since $\SiOtwo$ dust aggregates form new bonds when they are compressed.  Also, it is ideal for low velocity collisions since we do not consider vaporisation and melting and such processes are not as important in low velocity collisions.  Our model cannot simulate dust aggregate collisions that are faster than the sound speed in $\SiOtwo$ \citep[$\approx 30$~m/s;][]{Blum_Wurm_review} because at such high velocities, shocks will occur resulting in heating.  In this case, our porosity model breaks down as it assumes an isothermal process.

\section{Simulations}
\label{sec:sim}

\begin{table*}
\centering
  {\scriptsize
\begin{tabular}{llllll}
    \hline
    Simulation & Filling & $R_t$ & $R_p$ & $\vc$ & Growth?\\
    name & factor & [cm] & [cm] & [m/s] & \\
   \hline
    \hline
    10cm-2cm-35\%-20m/s & 35\% & 10.0 & 2.0 & 20 & Yes\\
    10cm-2cm-35\%-27.5m/s & 35\% & 10.0 & 2.0 & 27.5 & Yes\\
    10cm-3cm-35\%-20m/s & 35\% & 10.0 & 3.0 & 20 & Yes\\
    10cm-3cm-35\%-21m/s & 35\% & 10.0 & 3.0 & 21 & Yes\\
    10cm-3cm-35\%-22m/s & 35\% & 10.0 & 3.0 & 22 & No\\
    10cm-3cm-35\%-25m/s & 35\% & 10.0 & 3.0 & 25 & No\\
    10cm-4cm-35\%-15m/s & 35\% & 10.0 & 4.0 & 15 & Yes\\
    10cm-4cm-35\%-17.5m/s & 35\% & 10.0 & 4.0 & 17.5 & Yes\\
    10cm-4cm-35\%-18.5m/s & 35\% & 10.0 & 4.0 & 18.5 & Yes\\
    10cm-4cm-35\%-19.5m/s & 35\% & 10.0 & 4.0 & 19.5 & No\\
    10cm-4cm-35\%-20m/s & 35\% & 10.0 & 4.0 & 20 & No\\
    10cm-5cm-35\%-10m/s & 35\% & 10.0 & 5.0 & 10 & Yes\\
    10cm-5cm-35\%-15m/s & 35\% & 10.0 & 5.0 & 15 & Yes\\
    10cm-5cm-35\%-17.5m/s & 35\% & 10.0 & 5.0 & 17.5 & No\\
    10cm-5cm-35\%-20m/s & 35\% & 10.0 & 5.0 & 20 & No\\
    10cm-6cm-35\%-12m/s (Reference) & 35\% & 10.0 & 6.0 & 12.0 & Yes\\
    \emph{10cm-6cm-35\%-12.5m/s (Reference) } & \emph{35\%} & \emph{10.0} & \emph{6.0} & \emph{12.5} & \emph{No}\\
    10cm-7cm-35\%-10m/s & 35\% & 10.0 & 7.0 & 10 & Yes\\
    10cm-7cm-35\%-11.5m/s & 35\% & 10 & 7.0 & 11.5 & No\\
    10cm-8cm-35\%-10m/s & 35\% & 10.0 & 8.0 & 10 & Yes\\
    10cm-8cm-35\%-12.5m/s & 35\% & 10.0 & 8.0 & 12.5 & No\\
    10cm-9cm-35\%-7.5m/s & 35\% & 10.0 & 9.0 & 7.5 & Yes\\
    10cm-9cm-35\%-10m/s & 35\% & 10.0 & 9.0 & 10 & No\\
    {\bf 10cm-10cm-35\%-0.1m/s} & {\bf 35\%} & {\bf 10.0} & {\bf 10.0} & {\bf 0.1} & {\bf Yes}\\
    {\bf 10cm-10cm-35\%-0.2m/s} & {\bf 35\%} & {\bf 10.0} & {\bf 10.0} & {\bf 0.2} & {\bf Yes}\\
    {\bf 10cm-10cm-35\%-1m/s} & {\bf 35\%} & {\bf 10.0} & {\bf 10.0} & {\bf 1} & {\bf Yes}\\
    10cm-10cm-35\%-2m/s & 35\% & 10.0 & 10.0 & 2 & Yes \\
    10cm-10cm-35\%-2.2m/s & 35\% & 10.0 & 10.0 & 2.2 & Yes \\
    10cm-10cm-35\%-2.5m/s & 35\% & 10.0 & 10.0 & 2.5 & Yes \\
    10cm-10cm-35\%-4m/s & 35\% & 10.0 & 10.0 & 4 & Yes \\
    10cm-10cm-35\%-5m/s & 35\% & 10.0 & 10.0 & 5 & Yes \\
    10cm-10cm-35\%-6m/s & 35\% & 10.0 & 10.0 & 6 & Yes \\
    10cm-10cm-35\%-7.5m/s & 35\% & 10.0 & 10.0 & 7.5 & Yes \\
    10cm-10cm-35\%-10m/s & 35\% & 10.0 & 10.0 & 10 & No \\
    10cm-10cm-35\%-12.5m/s & 35\% & 10.0 & 10.0 & 12.5 & No \\
    10cm-10cm-35\%-17.5m/s & 35\% & 10.0 & 10.0 & 17.5 & No \\
    10cm-10cm-35\%-20m/s & 35\% & 10.0 & 10.0 & 20 & No \\
   \hline
    10cm-3cm-25\%-17.5m/s & 25\% & 10.0 & 3.0 & 17.5 & Yes\\
    10cm-3cm-25\%-22.5m/s & 25\% & 10.0 & 3.0 & 22.5 & No\\
    10cm-4cm-25\%-12.5m/s & 25\% & 10.0 & 4.0 & 12.5 & Yes\\
    10cm-4cm-25\%-15m/s & 25\% & 10.0 & 4.0 & 15 & No\\
    10cm-5cm-25\%-10m/s & 25\% & 10.0 & 5.0 & 10 & Yes\\
    10cm-5cm-25\%-12.5m/s & 25\% & 10.0 & 5.0 & 12.5 & No\\
    10cm-6cm-25\%-10m/s & 25\% & 10.0 & 6.0 & 10 & Yes\\
    10cm-6cm-25\%-11m/s & 25\% & 10.0 & 6.0 & 11 & No\\
    10cm-6cm-25\%-12m/s & 25\% & 10.0 & 6.0 & 12 & No\\
    10cm-6cm-25\%-12.5m/s & 25\% & 10.0 & 6.0 & 12.5 & No\\
    10cm-7cm-25\%-7.5m/s & 25\% & 10.0 & 7.0 & 7.5 & Yes\\
    10cm-7cm-25\%-10m/s & 25\% & 10.0 & 7.0 & 10 & No\\
    10cm-8cm-25\%-7.5m/s & 25\% & 10.0 & 8.0 & 7.5 & Yes\\
    10cm-8cm-25\%-10m/s & 25\% & 10.0 & 8.0 & 10 & No\\
    10cm-9cm-25\%-7.5m/s & 25\% & 10.0 & 9.0 & 7.5 & Yes\\
    10cm-9cm-25\%-10m/s & 25\% & 10.0 & 9.0 & 10 & No\\
    10cm-10cm-25\%-7.5m/s & 25\% & 10.0 & 10.0 & 7.5 & Yes\\
    10cm-10cm-25\%-9m/s & 25\% & 10.0 & 10.0 & 9 & No\\
   \hline
    10cm-2cm-15\%-17.5m/s & 15\% & 10.0 & 2.0 & 17.5 & Yes\\
    10cm-2cm-15\%-20m/s & 15\% & 10.0 & 2.0 & 20 & Yes\\
    10cm-3cm-15\%-10m/s & 15\% & 10.0 & 3.0 & 10 & Yes\\
    10cm-3cm-15\%-12m/s & 15\% & 10.0 & 3.0 & 12 & Yes\\
    10cm-3cm-15\%-13m/s & 15\% & 10.0 & 3.0 & 13 & Yes\\
    10cm-3cm-15\%-15m/s & 15\% & 10.0 & 3.0 & 15 & No\\
    10cm-4cm-15\%-7.5m/s & 15\% & 10.0 & 4.0 & 7.5 & Yes\\
    10cm-4cm-15\%-10m/s & 15\% & 10.0 & 4.0 & 10 & No\\
    10cm-5cm-15\%-7.5m/s & 15\% & 10.0 & 5.0 & 7.5 & Yes\\
    10cm-5cm-15\%-9m/s & 15\% & 10.0 & 5.0 & 9 & No\\
    10cm-5cm-15\%-10m/s & 15\% & 10.0 & 5.0 & 10 & No\\
    10cm-6cm-15\%-6m/s & 15\% & 10.0 & 6.0 & 6 & Yes\\
    10cm-6cm-15\%-7.5m/s & 15\% & 10.0 & 6.0 & 7.5 & No\\
    10cm-7cm-15\%-6m/s & 15\% & 10.0 & 7.0 & 6 & Yes\\ 
    10cm-7cm-15\%-7m/s & 15\% & 10.0 & 7.0 & 7 & No\\
    10cm-8cm-15\%-6m/s & 15\% & 10.0 & 8.0 & 6 & Yes\\
    10cm-8cm-15\%-7.5m/s & 15\% & 10.0 & 8.0 & 7.5 & No\\
    10cm-9cm-15\%-6m/s & 15\% & 10.0 & 9.0 & 6 & Yes\\
    10cm-9cm-15\%-7m/s & 15\% & 10.0 & 9.0 & 7 & No\\
    10cm-10cm-15\%-6m/s & 15\% & 10.0 & 10.0 & 6 & Yes\\
    10cm-10cm-15\%-7.5m/s & 15\% & 10.0 & 10.0 & 7.5 & No\\
    \hline
  \end{tabular}
}
  \caption{Table showing the simulations carried out to investigate the effect that the projectile size (or equivalently, mass) has on the fragmentation velocity for aggregates with filling factors of 35\% (top panel), 25\% (middle panel) and 15\% (bottom panel), and key growth results.  We also include the relevant simulation carried out in Section 5 of \citet{Four_population} (italicised).  The simulations which we loosely define to be roughly neutral (i.e. bouncing occurs), defined by equations~\ref{eq:neutral_ml} and~\ref{eq:neutral_ms}, are highlighted in bold.}
 \label{tab:sim_mass}
\end{table*}

\begin{table*}
\centering
  {\small
\begin{tabular}{llllll}
    \hline
    Simulation & Filling & $R_t$ & $R_p$ & $\vc$ & Growth?\\
    name & factor & [cm] & [cm] & [m/s] & \\
   \hline
    \hline
    6cm-6cm-35\%-7.5m/s & 35\% & 6.0 & 6.0 & 7.5 & Yes\\
    6cm-6cm-35\%-10m/s & 35\% & 6.0 & 6.0 & 10.0 & No\\
    6cm-6cm-35\%-12.5m/s & 35\% & 6.0 & 6.0 & 12.5 & No\\
    8cm-6cm-35\%-10m/s & 35\% & 8.0 & 6.0 & 10.0 & Yes\\
    8cm-6cm-35\%-11m/s & 35\% & 8.0 & 6.0 & 11.0 & No\\
    8cm-6cm-35\%-12.5m/s & 35\% & 8.0 & 6.0 & 12.5 & No\\
    10cm-6cm-35\%-12m/s (Reference) & 35\% & 10.0 & 6.0 & 12.0 & Yes\\
    \emph{10cm-6cm-35\%-12.5m/s (Reference) } & \emph{35\%} & \emph{10.0} & \emph{6.0} & \emph{12.5} & \emph{No}\\
  \hline
    5cm-3cm-35\%-12.5m/s & 35\% & 5.0 & 3.0 & 12.5 & Yes\\
    5cm-3cm-35\%-15m/s & 35\% & 5.0 & 3.0 & 15.0 & Yes\\
    5cm-3cm-35\%-20m/s & 35\% & 5.0 & 3.0 & 20.0 & No\\
    8.3cm-5cm-35\%-14m/s & 35\% & 8.3 & 5.0 & 14.0 & Yes\\
    8.3cm-5cm-35\%-15m/s & 35\% & 8.3 & 5.0 & 15.0 & No\\
    10cm-6cm-35\%-12m/s (Reference) & 35\% & 10.0 & 6.0 & 12.0 & Yes\\
   \emph{10cm-6cm-35\%-12.5m/s (Reference) } & \emph{35\%} & \emph{10.0} & \emph{6.0} & \emph{12.5} & \emph{No}\\
   15cm-9cm-35\%-10m/s & 35\% & 15.0 & 9.0 & 10.0 & Yes\\
   15cm-9cm-35\%-11.5m/s & 35\% & 15.0 & 9.0 & 11.5 & No\\
   \hline
  \end{tabular}
}
  \caption{Table showing the simulations carried out to investigate the impact of maintaining the same projectile mass but varying the target mass (top panel) and how changing the aggregate masses (while maintaining the same mass ratio) affects the fragmentation velocity (bottom panel), as well as the key growth results.  We also include the relevant simulation carried out in Section 5 of \citet{Four_population} (italicised).}
 \label{tab:sim_mass_ratio}
\end{table*}

\begin{table*}
\centering
  {\scriptsize
\begin{tabular}{llllll}
    \hline
    Simulation & Filling & $R_t$ & $R_p$ & $\vc$ & Growth?\\
    name & factor & [cm] & [cm] & [m/s] & \\
    \hline
    \hline
    10cm-4cm-15\%-7.5m/s & 15\% & 10.0 & 4.0 & 7.5 & Yes\\
    10cm-4cm-15\%-10m/s & 15\% & 10.0 & 4.0 & 10 & No\\
    10cm-4cm-20\%-12m/s & 20\% & 10.0 & 4.0 & 12 & Yes\\
    10cm-4cm-20\%-14m/s & 20\% & 10.0 & 4.0 & 14 & No\\
    10cm-4cm-25\%-12.5m/s & 25\% & 10.0 & 4.0 & 12.5 & Yes\\
    10cm-4cm-25\%-15m/s & 25\% & 10.0 & 4.0 & 15 & No\\
    10cm-4cm-30\%-15m/s & 30\% & 10.0 & 4.0 & 15 & Yes\\
    10cm-4cm-30\%-17.5m/s & 30\% & 10.0 & 4.0 & 17.5 & No\\
    10cm-4cm-30\%-20m/s & 30\% & 10.0 & 4.0 & 20 & No\\
    10cm-4cm-35\%-18.5m/s & 35\% & 10.0 & 4.0 & 18.5 & Yes\\
    10cm-4cm-35\%-19.5m/s & 35\% & 10.0 & 4.0 & 19.5 & No\\
    10cm-4cm-35\%-20m/s & 35\% & 10.0 & 4.0 & 20 & No\\
    {\bf 10cm-4cm-40\%-2.0m/s} & {\bf 40\%} & {\bf 10.0} & {\bf 4.0} & {\bf 2.0} & {\bf Yes}\\
    10cm-4cm-40\%-3.0m/s & 40\% & 10.0 & 4.0 & 3.0 & No\\
    \hline
    10cm-6cm-15\%-0.1m/s & 15\% & 10.0 & 6.0 & 0.1 & Yes\\
    10cm-6cm-15\%-0.2m/s & 15\% & 10.0 & 6.0 & 0.2 & Yes\\
    10cm-6cm-15\%-0.3m/s & 15\% & 10.0 & 6.0 & 0.3 & Yes\\
    10cm-6cm-15\%-0.5m/s & 15\% & 10.0 & 6.0 & 0.5 & Yes\\
    10cm-6cm-15\%-1m/s & 15\% & 10.0 & 6.0 & 1 & Yes\\
    10cm-6cm-15\%-2m/s & 15\% & 10.0 & 6.0 & 2 & Yes\\
    10cm-6cm-15\%-2.5m/s & 15\% & 10.0 & 6.0 & 2.5 & Yes\\
    10cm-6cm-15\%-4m/s & 15\% & 10.0 & 6.0 & 4.0 & Yes\\
    10cm-6cm-15\%-5m/s & 15\% & 10.0 & 6.0 & 5.0 & Yes\\
    10cm-6cm-15\%-6m/s & 15\% & 10.0 & 6.0 & 6.0 & Yes\\
    10cm-6cm-15\%-7.5m/s & 15\% & 10.0 & 6.0 & 7.5 & No\\
    10cm-6cm-15\%-10m/s & 15\% & 10.0 & 6.0 & 10.0 & No\\
    10cm-6cm-15\%-12.5m/s & 15\% & 10.0 & 6.0 & 12.5 & No\\
    10cm-6cm-15\%-17.5m/s & 15\% & 10.0 & 6.0 & 17.5 & No\\
    10cm-6cm-15\%-20m/s & 15\% & 10.0 & 6.0 & 20.0 & No\\
    10cm-6cm-20\%-8m/s & 20\% & 10.0 & 6.0 & 8 & Yes\\
    10cm-6cm-20\%-9m/s & 20\% & 10.0 & 6.0 & 9 & No\\
    10cm-6cm-25\%-10m/s & 25\% & 10.0 & 6.0 & 10 & Yes\\
    10cm-6cm-25\%-11m/s & 25\% & 10.0 & 6.0 & 11 & No\\
    10cm-6cm-25\%-12m/s & 25\% & 10.0 & 6.0 & 12 & No\\
    10cm-6cm-25\%-12.5m/s & 25\% & 10.0 & 6.0 & 12.5 & No\\
    10cm-6cm-30\%-10m/s & 30\% & 10.0 & 6.0 & 10 & Yes\\
    10cm-6cm-30\%-11.5m/s & 30\% & 10.0 & 6.0 & 11.5 & No\\
    10cm-6cm-30\%-12.5m/s & 30\% & 10.0 & 6.0 & 12.5 & No\\
    10cm-6cm-35\%-12m/s (Reference) & 35\% & 10.0 & 6.0 & 12.0 & Yes\\
    \emph{10cm-6cm-35\%-12.5m/s (Reference) } & \emph{35\%} & \emph{10.0} & \emph{6.0} & \emph{12.5} & \emph{No}\\
   10cm-6cm-37\%-11.5m/s & 37\% & 10.0 & 6.0 & 11.5 & Yes\\
    {\bf 10cm-6cm-39\%-2.5m/s} & {\bf 39\%} & {\bf 10.0} & {\bf 6.0} & {\bf 2.5} & {\bf No}\\
    10cm-6cm-39\%-11.5m/s & 39\% & 10.0 & 6.0 & 11.5 & No\\
    {\bf 10cm-6cm-40\%-2m/s} & {\bf 40\%} & {\bf 10.0} & {\bf 6.0} & {\bf 2} & {\bf Yes}\\
    {\bf 10cm-6cm-40\%-2.5m/s} & {\bf 40\%} & {\bf 10.0} & {\bf 6.0} & {\bf 2.5} & {\bf No}\\
    10cm-6cm-40\%-5m/s & 40\% & 10.0 & 6.0 & 5 & No\\
    10cm-6cm-40\%-7.5m/s & 40\% & 10.0 & 6.0 & 7.5 & No\\
    {\bf 10cm-6cm-45\%-1m/s} & {\bf 45\%} & {\bf 10.0} & {\bf 6.0} & {\bf 1} & {\bf No}\\
    {\bf 10cm-6cm-45\%-2m/s} & {\bf 45\%} & {\bf 10.0} & {\bf 6.0} & {\bf 2} & {\bf No}\\
    10cm-6cm-45\%-2.5m/s & 45\% & 10.0 & 6.0 & 2.5 & No\\
    10cm-6cm-45\%-5m/s & 45\% & 10.0 & 6.0 & 5 & No\\
    10cm-6cm-45\%-7.5m/s & 45\% & 10.0 & 6.0 & 7.5 & No\\
   10cm-6cm-50\%-2m/s & 50\% & 10.0 & 6.0 & 2 & No\\
    10cm-6cm-50\%-2.5m/s & 50\% & 10.0 & 6.0 & 2.5 & No\\
    10cm-6cm-50\%-3m/s & 50\% & 10.0 & 6.0 & 3 & No\\
    10cm-6cm-50\%-4m/s & 50\% & 10.0 & 6.0 & 4 & No\\
    {\bf 10cm-6cm-55\%-2m/s} & {\bf 55\%} & {\bf 10.0} & {\bf 6.0} & {\bf 2} & {\bf No}\\
    10cm-6cm-55\%-2.5m/s & 55\% & 10.0 & 6.0 & 2.5 & No\\
    \hline
    10cm-8cm-15\%-6m/s & 15\% & 10.0 & 8.0 & 6 & Yes\\
    10cm-8cm-15\%-7.5m/s & 15\% & 10.0 & 8.0 & 7.5 & No\\
    10cm-8cm-20\%-7m/s & 20\% & 10.0 & 8.0 & 7 & Yes\\
    10cm-8cm-20\%-9m/s & 20\% & 10.0 & 8.0 & 9 & No\\
    10cm-8cm-25\%-7.5m/s & 25\% & 10.0 & 8.0 & 7.5 & Yes\\
    10cm-8cm-25\%-10m/s & 25\% & 10.0 & 8.0 & 10 & No\\
    10cm-8cm-30\%-9m/s & 30\% & 10.0 & 8.0 & 9 & Yes\\
    10cm-8cm-30\%-10m/s & 30\% & 10.0 & 8.0 & 10 & No\\
    10cm-8cm-35\%-10m/s & 35\% & 10.0 & 8.0 & 10 & Yes\\
    10cm-8cm-35\%-12.5m/s & 35\% & 10.0 & 8.0 & 12.5 & No\\
    {\bf 10cm-8cm-40\%-2.0m/s} & {\bf 40\%} & {\bf 10.0} & {\bf 8.0} & {\bf 2.0} & {\bf Yes}\\
    {\bf 10cm-8cm-40\%-3.0m/s} & {\bf 40\%} & {\bf 10.0} & {\bf 8.0} & {\bf 3.0} & {\bf No}\\
    \hline
  \end{tabular}
}
  \caption{Table showing the simulations carried out to investigate the effect that the aggregate porosity has on the fragmentation velocity for projectile radii, $R_{\rm p} = 4$ (top panel), 6 (middle panel) and 8~cm (bottom panel), as well as the key growth results.  The porosity of the target and projectile are identical in each simulation.  We also include the relevant simulation carried out in Section 5 of \citet{Four_population} (italicised).  The simulations which we loosely define to be roughly neutral (i.e. bouncing occurs), defined by equations~\ref{eq:neutral_ml} and~\ref{eq:neutral_ms}, are highlighted in bold.}
   \label{tab:sim_porosity}
\end{table*}

We carry out three-dimensional Smoothed Particle Hydrodynamics (SPH) simulations of head-on collisions between $\rm SiO_2$ dust aggregates with radii between $2-15$~cm with the target radius, $\Rt$, larger than or equal to that of the projectile, $\Rp$.

The Reference case that we compare all our simulations to has the same setup as that in Section 5 of \cite{Four_population}: a projectile with a radius, $\Rp = 6 \rm cm$, colliding with a stationary target with radius, $\Rt = 10 \rm cm$.  These aggregates are modelled using 238,238 and 51,477 SPH particles (note that this varies for different sized aggregates since the number of SPH particles per unit volume is kept constant to maintain the same spatial resolution in all the simulations).

The dust aggregates in the Reference case both have initial filling factors of $\phi = 35\%$, such that their masses are 0.63 and 2.93~kg for the projectile and target, respectively.  \cite{Four_population} found that the fragmentation velocity with this setup was $\approx 12$~m/s.

We firstly investigate how the size ratio (or equivalently, the mass ratio) of the aggregates affects the fragmentation velocities.  Table~\ref{tab:sim_mass} outlines the simulations and the key growth results.  We vary the radius of the projectile to between $20-100\%$ of the target radius, while maintaining the target radius of 10cm and the initial filling factors of the aggregates at $\phi = 35\%$, and determine how the threshold velocity for fragmentation (defined to occur when the mass of the largest aggregate following the collision is smaller than the initial mass of the target) varies as the projectile size changes.  We then carry out another two suites of simulations, repeating this exercise for aggregates with filling factors, $\phi = 15\%$ and $25\%$.  In addition, for aggregates with $\phi = 35\%$ we investigate the effect that changing the target mass has on the fragmentation velocity (while maintaining the projectile as in the Reference case) as well as the effect that changing both the target and projectile size in proportion such that the mass ratio remains constant (see Table~\ref{tab:sim_mass_ratio} for details).

Secondly, we investigate how the porosity of the aggregates affect the fragmentation velocity.  Using the same setup of the aggregate radii as the Reference case (6cm projectile impacting a 10cm target), we vary the filling factor of the aggregates (between 15-55\%) and determine the effects on the fragmentation velocity.  Note that in any one simulation, the initial filling factors of the target and projectile are identical.  Table~\ref{tab:sim_porosity} outlines the simulations performed and the key results on the collisional outcome.

We find that if the aggregates break apart they do so within the first 0.7 seconds.  We therefore run each of these simulations for a physical time of one second or until the aggregates have fragmented (defined to be when the mass of the largest fragment is smaller than the initial target mass).

The SPH particles are set up using a simple cubic lattice and by aligning the SPH particles such that they are parallel to the axis along which the collision occurs.  We also carry out some numerical tests to show that the results for the threshold velocity are similar even when the SPH particles are not all aligned with the collision axis and when a different lattice type is used (see Appendix~\ref{appendix}).

\section{Results}
\label{sec:results}

The simulations have been analysed to classify their collisional outcomes as positive or negative growth (i.e. sticking or fragmentation, respectively) if the mass of the largest aggregate after the collision is smaller or larger, respectively, than the initial mass of the target.  Within the context of the four-population model presented by \cite{Four_population}, our focus is primarily on the population representing the most massive object after the collision.  The fragmentation threshold is defined to be half way between the lowest velocity at which negative growth occurs and the highest velocity at which positive growth occurs.

Note that strictly speaking, in between the regions where postive and negative growth occur, a region exists where growth is neither positive or negative, i.e. neutral.  In our simulations a strictly neutral regime will never exist as some mass is always added or removed from either of the aggregates.  Therefore, we also define a neutral regime when the masses of both the largest and second largest fragments differ from the initial target and projectile masses, respectively, by less than 6\% of the initial projectile mass, i.e. when

\begin{equation}
\mt - 0.06 \mproj \leq m_{\rm 1} \leq \mt + 0.06 \mproj
\label{eq:neutral_ml}
\end{equation}
and

\begin{equation}
\mt - 0.06 \mproj \leq m_{\rm 2} \leq \mt + 0.06 \mproj
\label{eq:neutral_ms}
\end{equation}
where $m_{\rm 1}$ and $m_{\rm 2}$ are the massive and second massive objects after the collision, respectively.  Note that this definition includes the scenaria where the target and projectile may either gain or lose a small amount of their mass.  While the definition has been motivated by experimental results from \cite{Beitz_bouncing}, we point out that this is an arbitrary quantitative definition that we use purely for the sake of illustrating the results.

\subsection{Aggregate size/mass}
\label{sec:mass}

\subsubsection{Projectile size/mass}
\label{sec:mproj}

\begin{figure*}
\centering
  \includegraphics[width=1.0\columnwidth]{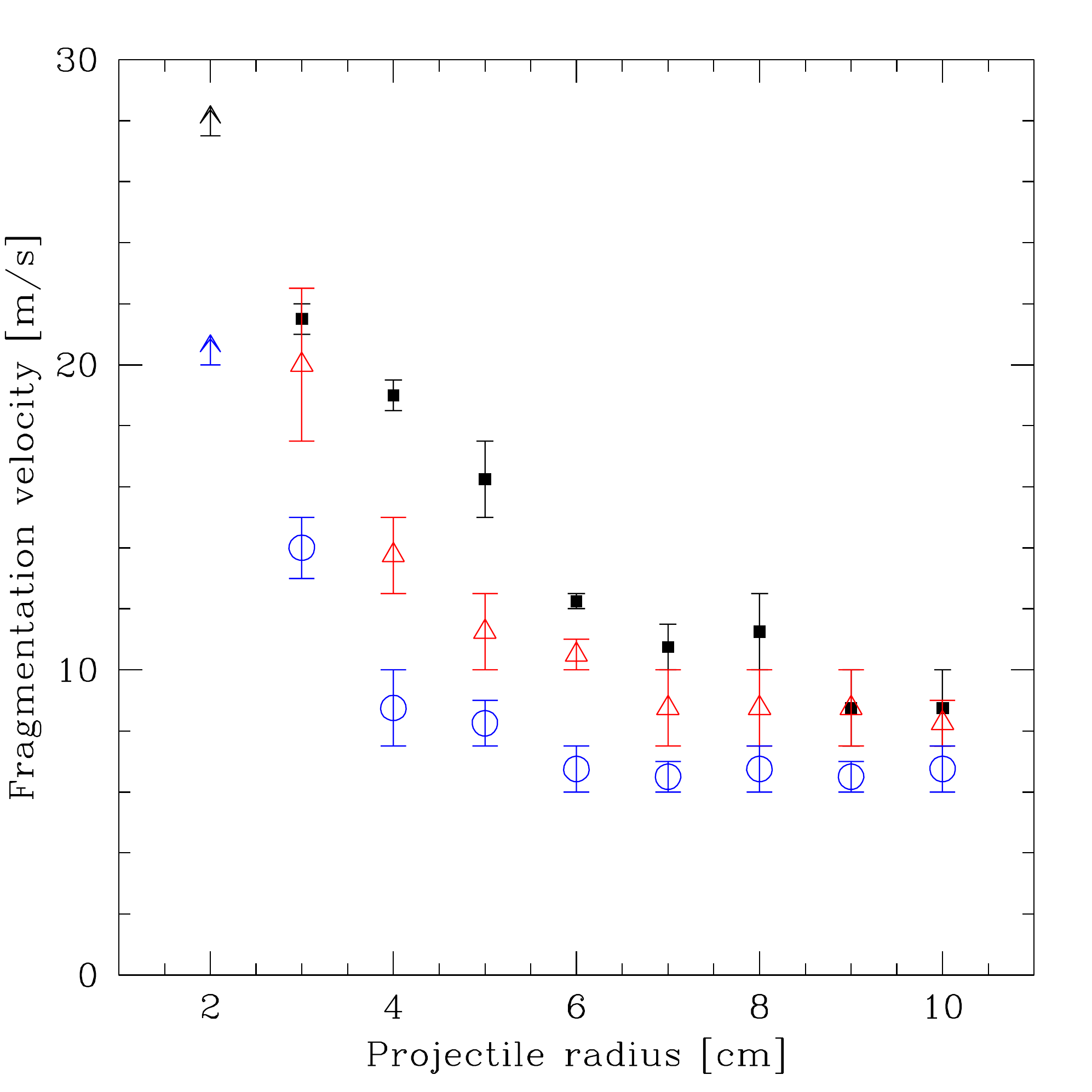}
  \includegraphics[width=1.0\columnwidth]{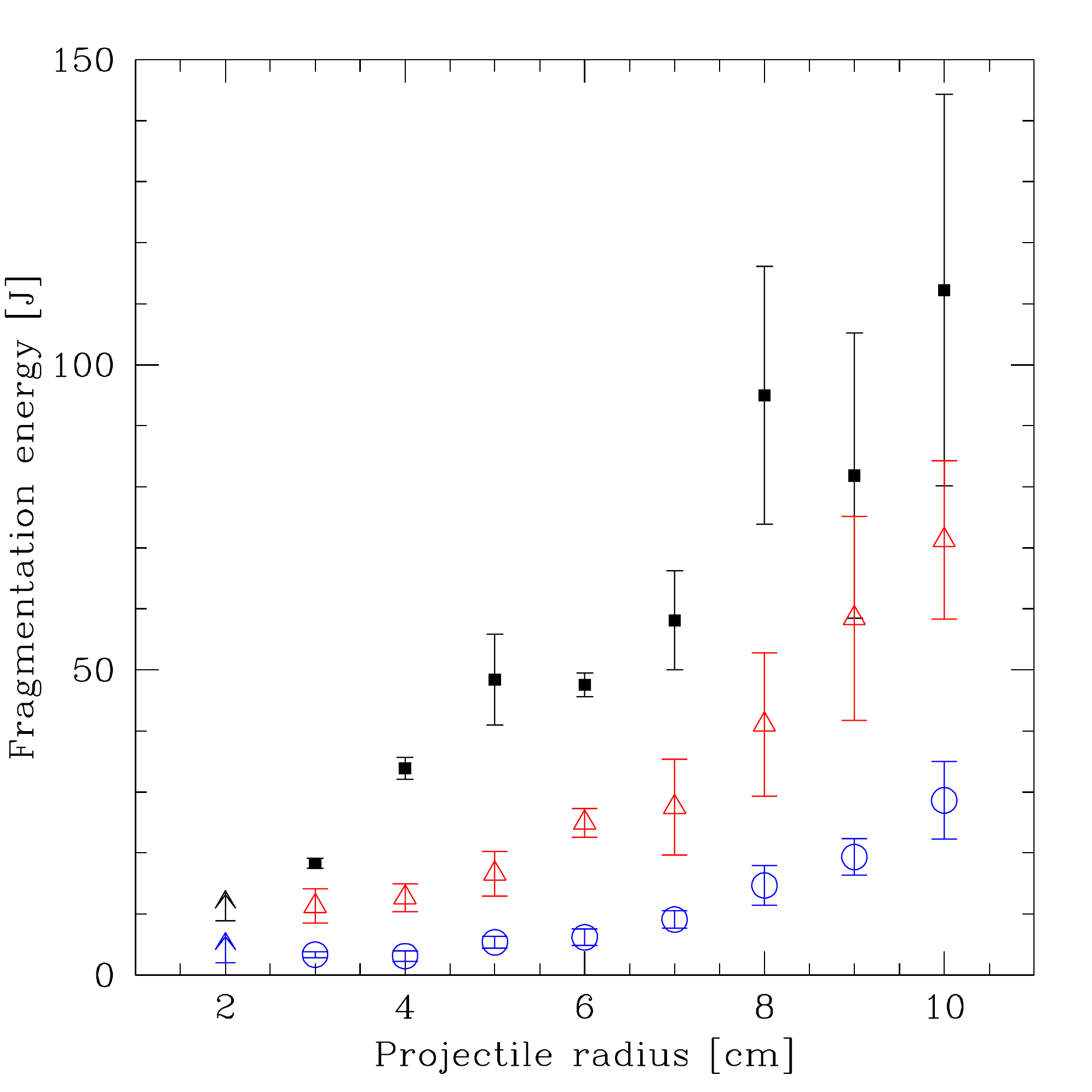}
  \caption{Graph showing the fragmentation velocity (left panel) and fragmentation energy in the centre of mass frame (right panel) against projectile radius for simulations with filling factors of 35\% (black squares), 25\% (red triangles) and 15\% (blue circles).  The error bars show the upper and lower limits around the threshold velocities and energies, obtained from the simulation results.  Where only the lower limit exists, it is shown using an upwards arrow.  The left panel shows that the fragmentation velocity decreases as the projectile size increases).  In addition, as the projectile size increases the energy required to break the aggregate apart also increases (right panel).  The target radius is $\Rt = 10$~cm.}
\label{fig:vth_size}
\end{figure*}

\begin{figure*}
\centering
\includegraphics[width=2.0\columnwidth]{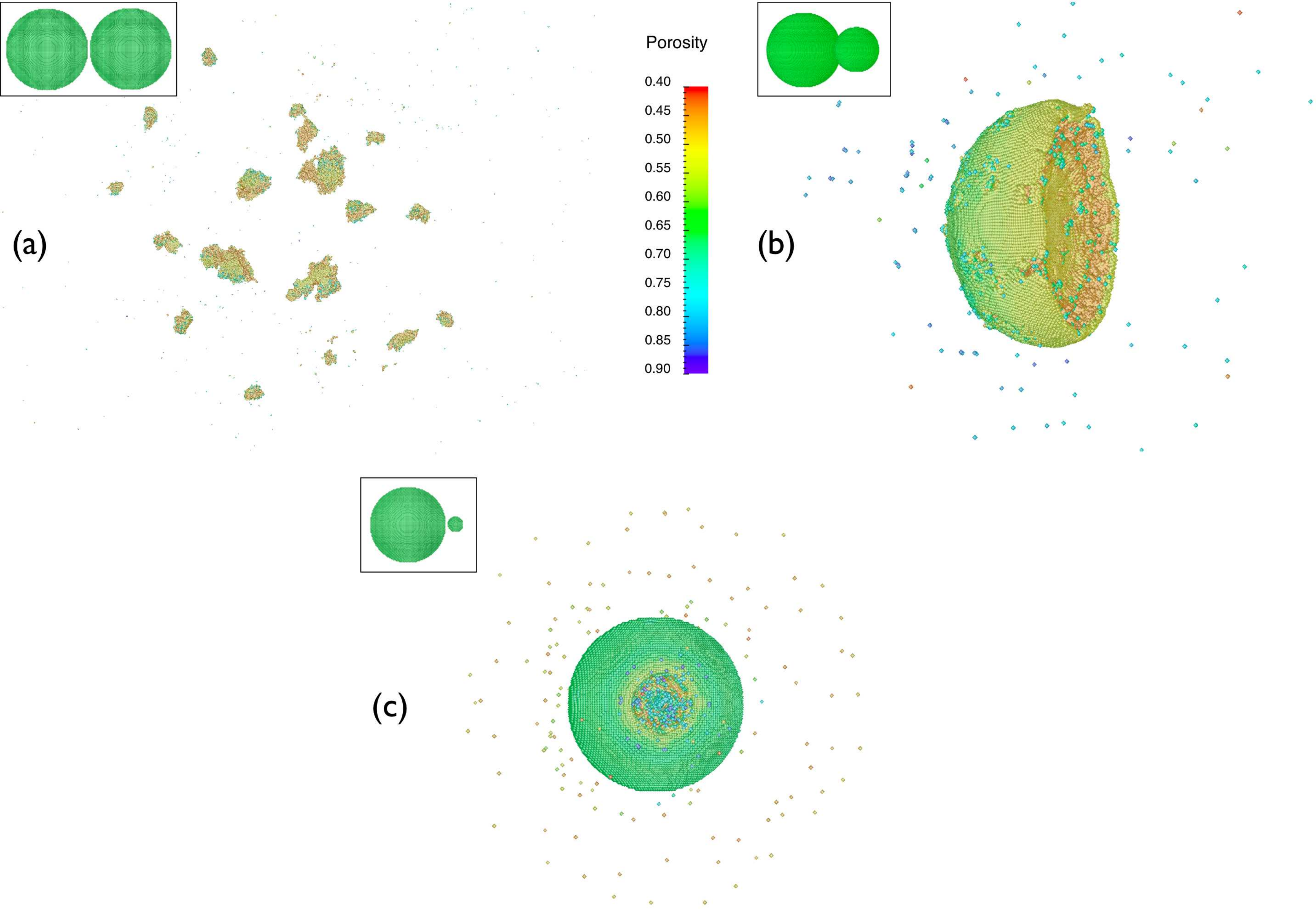}
  \caption{Porosity rendered images showing the outcome of collisions between a target with radius, $\Rt = 10$~cm and projectiles with radii (a) $\Rp = 10$~cm, with collision velocity $\vc = 10$m/s, (b) $\Rp = 6$cm with $\vc = 10$m/s and (c) $\Rp = 2$~cm with $\vc = 20$~m/s.  By decreasing the projectile size, the aggregates are able to withstand a higher collision velocity and do not fragment as easily.  The initial filling factor is $\phi = 35\%$.  The insets show the initial relative aggregate sizes.}
\label{fig:size_illus}
\end{figure*}

\begin{table}
\centering
\begin{tabular}{llll}
    \hline
    & $\phi = 0.15$ & $\phi = 0.25$ & $\phi = 0.35$\\
    \hline
    \hline
    $\eta_{\rm r}$ & $25 \pm 5$ & $34 \pm 5$ & $55 \pm 5$\\
    $\sigma_{\rm r}$ & $-0.7 \pm 0.1$ & $-0.65 \pm 0.08$ & $-0.82 \pm 0.06$\\
    \hline
    $\eta_{\rm m}$ & $26 \pm 6$ & $40 \pm 7$ & $73 \pm 9$\\
    $\sigma_{\rm m}$ & $-0.22 \pm 0.04$ & $-0.22 \pm 0.03$ & $-0.27 \pm 0.02$\\
    \hline
  \end{tabular}
 \caption{The fit parameters for equations~\ref{eq:fit_Rp} and~\ref{eq:fit_mp} showing how the threshold velocity for fragmentation depends on the projectile radius and mass, respectively, for filling factors, $\phi = 15\%$, $\phi = 25\%$ and $\phi = 35\%$.  The units of $\eta_{\rm r}$ and $\eta_{\rm m}$ are in cgs.}
 \label{tab:fits}
\end{table}

Table~\ref{tab:sim_mass} summarises the simulations carried out to understand how the fragmentation velocity varies with projectile size, as well as the key collisional outcomes.  Figure~\ref{fig:vth_size} (left panel) shows that as the projectile size increases the threshold velocity decreases and thus the smaller the projectile, the more likely that the outcome is growth.  This is because the larger structure of one aggregate relative to the other means that more energy can be absorbed in elastic loading and then dissipated via plastic deformation.  Therefore the more extreme mass ratio aggregates are more efficient at dissipating the collisional energy, resulting in a higher threshold for fragmentation.  In addition, Figure~\ref{fig:vth_size} (right panel) shows that the energy required to break an aggregate apart given by $\mproj v_{\rm th}^2 / 2$ increases with projectile size and therefore a single critical energy cannot be assumed for fragmentation.  These trends hold for all three filling factors considered.  Figure~\ref{fig:vth_size} (left panel) shows that the difference in velocity threshold is as much as a factor of $\approx 3$ (comparing simulations where the projectile is a similar size to the target and where the projectile and target are of very different sizes).  In particular, Figure~\ref{fig:vth_size} also shows that the porosity of the aggregate plays a very important role in determining the magnitude of the fragmentation velocity, with the effect of changing the porosity on the fragmentation velocity being more pronounced for smaller projectiles.

Figure~\ref{fig:size_illus} shows sample simulation results demonstrating that as the projectile radius is decreased from $\Rp = 10$~cm to 6cm (maintaining the same target radius of $\Rt = 10$~cm), the aggregates are able to withstand a collision velocity of 10m/s with the smaller projectile, whereas with the larger projectile the aggregates are shattered.  Figure~\ref{fig:size_illus} also shows that when the projectile radius is decreased further to $\Rp = 2$~cm, the aggregates withstand a higher collision velocity of 20m/s without breaking apart.  To estimate the dependence of the fragmentation velocity on the radius and mass of the projectiles we fit formulae of the form

\begin{equation}
v_{\rm th} = \eta_{\rm r} \Rp^{\sigma_{\rm r}}
\label{eq:fit_Rp}
\end{equation}
and

\begin{equation}
v_{\rm th} = \eta_{\rm m} \mproj^{\sigma_{\rm m}},
\label{eq:fit_mp}
\end{equation}
where $v_{\rm th}$ is the threshold velocity for fragmentation and $\eta_{\rm r}$, $\eta_{\rm m}$, $\sigma_{\rm r}$ and $\sigma_{\rm m}$ are constants to be determined and the subscripts 'r' and 'm' refer to 'radius' and 'mass', respectively.  Table~\ref{tab:fits} shows a summary of the fits found and shows that while the porosity of the aggregate clearly affects the dependence of the fragmentation velocity on the projectile size and masses, the dependence is broadly in the region $v_{\rm th} \propto \Rp^{\sigma_{\rm r}}$, where $\sigma_{\rm r}$ ranges from approximately -0.6 to -0.9, and $v_{\rm th} \propto \mproj^{\sigma_{\rm m}}$, where $\sigma_{\rm m}$ ranges from approximately -0.2 to -0.3.

\begin{figure}
\centering
  \includegraphics[width=1.0\columnwidth]{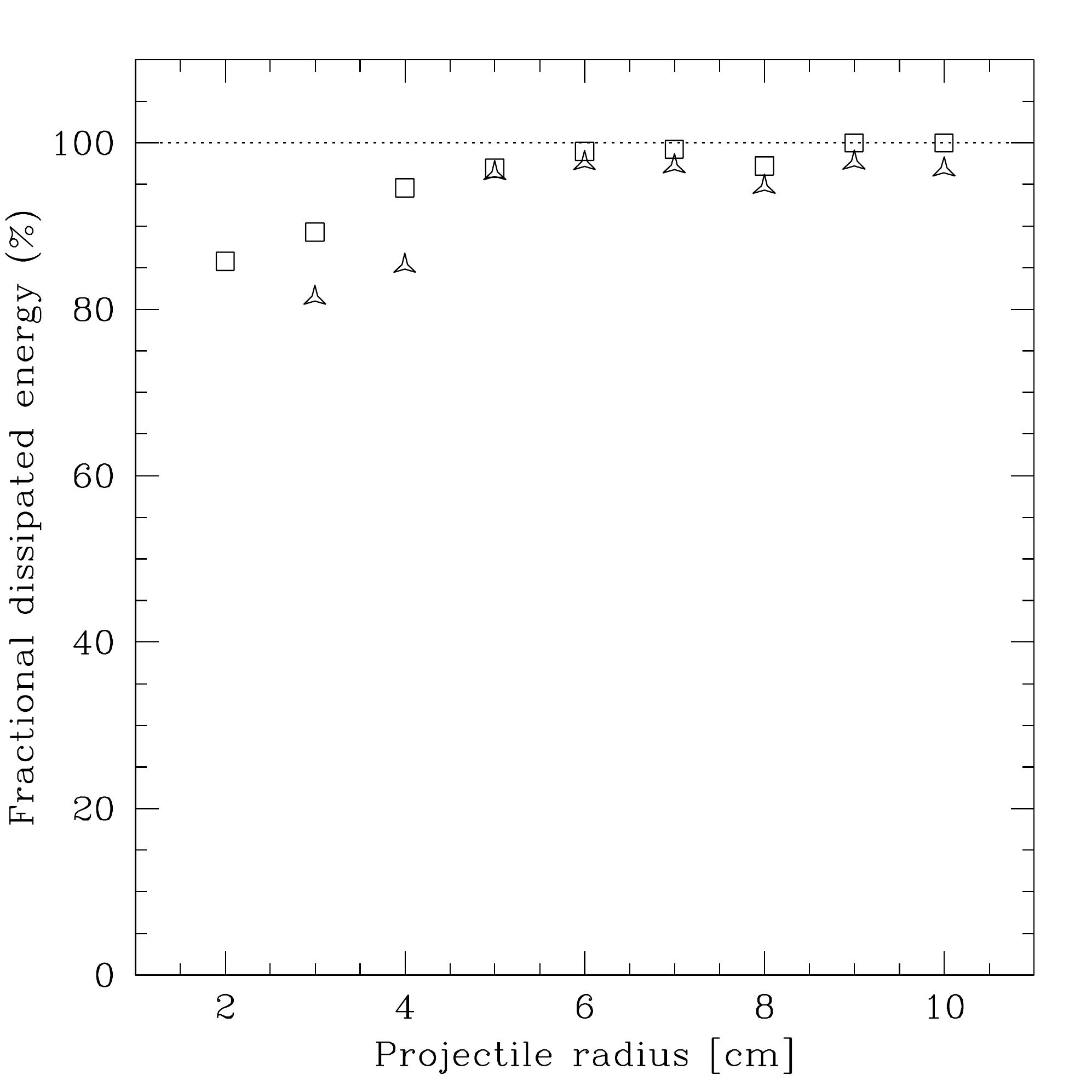}
  \caption{Graph showing the energy dissipated as a percentage of the maximum available energy in the collision against the projectile radius, for each of the simulations carried out that are closest to the threshold velocity that involve growth (squares) and fragmentation (triangles).  The dotted line marks where perfect sticking can occur (Equation~\ref{eq:Ediss}), i.e. the maximum amount of energy that can be dissipated.  The simulation data will always be at or below this maximum line.  The closer the data point is to 100\%, the closer it is to the threshold velocity.  The target radius is $\Rt = 10$~cm and the aggregate filling factors are 35\%.}
\label{fig:ediss_boundary_analytical}
\end{figure}

Figure~\ref{fig:ediss_boundary_analytical} shows the fraction of available energy that is dissipated for the simulations just either side of the fragmentation velocity for each radius, i.e. $\Ediss/\Edissmax$, expressed as a percentage, where $\Ediss$ is measured from the simulations by considering the kinetic energy of all the aggregates after the collision.  For perfect sticking 100\% of the available energy is dissipated (and this marks the boundary between the growth and fragmentation regions; Equation~\ref{eq:Ediss}), while for imperfect sticking or for fragmentation the dissipated energy will always be less than this (since in the fragmenting case, some of the energy will not have been dissipated and will instead go into kinetic energy of the fragments, while in the growth case not all the material will stick to the aggregate and will therefore go into the kinetic energy of the fragments that do not stick).  Since the simulations will never involve absolutely perfect sticking the results will always be below 100\%, though how close the data points are to this limit indicate how close the simulations are to the velocity threshold.  Since the simulations with projectile radii of 5cm and above are quite close to the fragmentation velocity, their fractional dissipated energies are close to 100\%.  We note that the data points for $\Rp=2$, 3 and 4~cm are much lower than 100\%, i.e. all the energy has not been used to stick the aggregates together.  This indicates that the simulations that have been carried out are not sufficiently close to the fragmentation velocities (e.g. for $\Rp = 2$~cm the energy dissipated is $\approx 85 \%$ of the energy available in the collision).  However, since the code that we use cannot be used for simulations involving collision velocities above $\approx 30$~m/s (see Section~\ref{sec:numerics}), we cannot run simulations that are closer to the perfect sticking case (or equivalently, the fragmentation threshold) for $\Rp = 2$~cm.  For the simulations involving 3~cm and 4~cm projectiles, we expect the fractional dissipated energy would move closer to 100\% if the choice of collision velocities was closer to the actual fragmentation velocity.  We stress that a comparison with the analytically expected values for the dissipated energy $\Edissmax$ (Equation~\ref{eq:Ediss}), can only be carried out in a regime where growth can occur.  In Section~\ref{sec:porosity} we show that regions of the parameter space exist where growth is not possible.

\subsubsection{Target size/mass}
\label{sec:mtarget}

\begin{figure}
\centering
  \includegraphics[width=1.0\columnwidth]{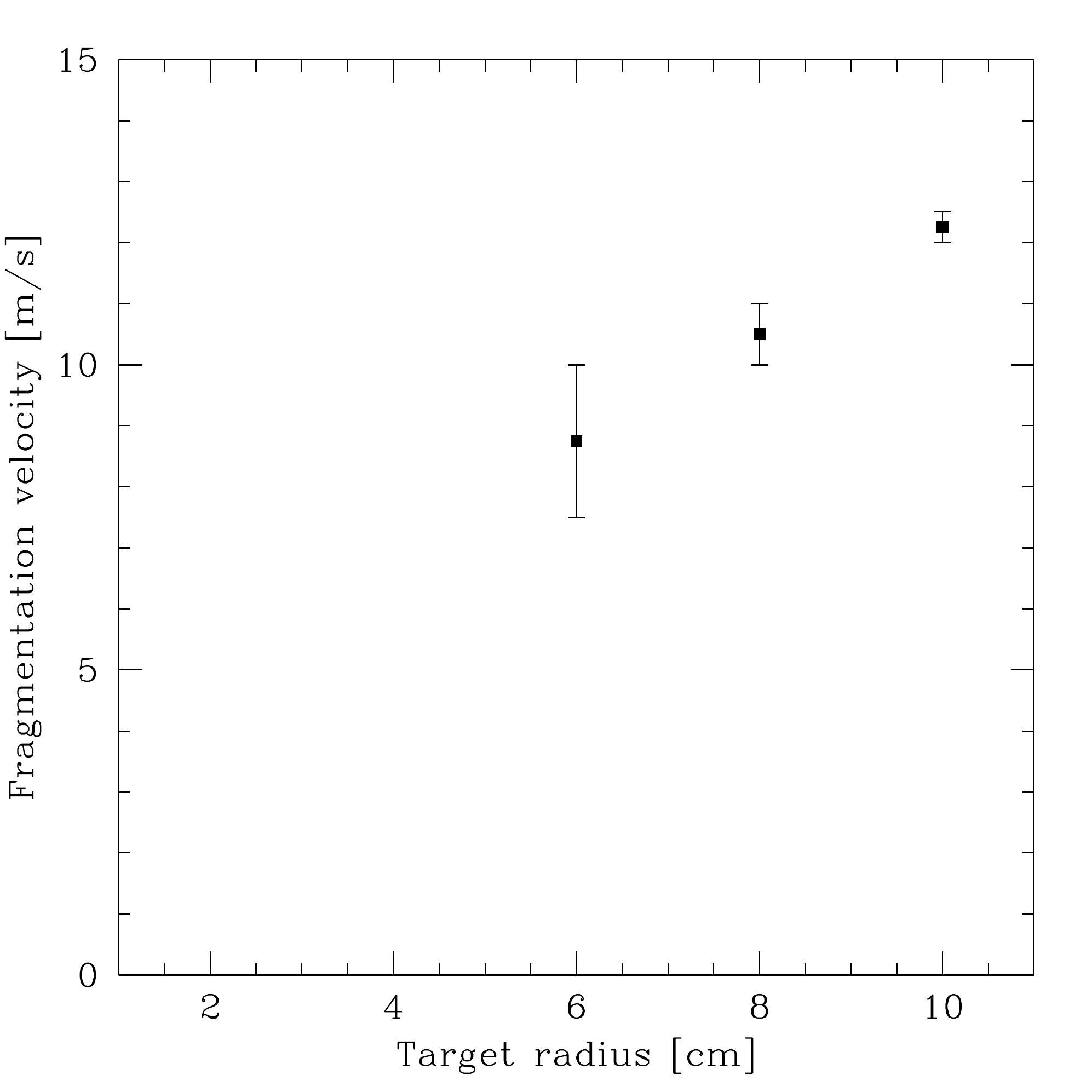}
  \caption{Graph showing the fragmentation velocity against target radius for simulations carried out with projectile radii, $R_{\rm p} = 6$cm, and filling factor, $\phi = 35\%$.  The error bars show the upper and lower limits around the threshold velocities, obtained from the simulation results.  The fragmentation velocity increases as the target size increases, consistent with the analytical theory.}
\label{fig:vth_tsize}
\end{figure}

\begin{figure}
\centering
  \includegraphics[width=1.0\columnwidth]{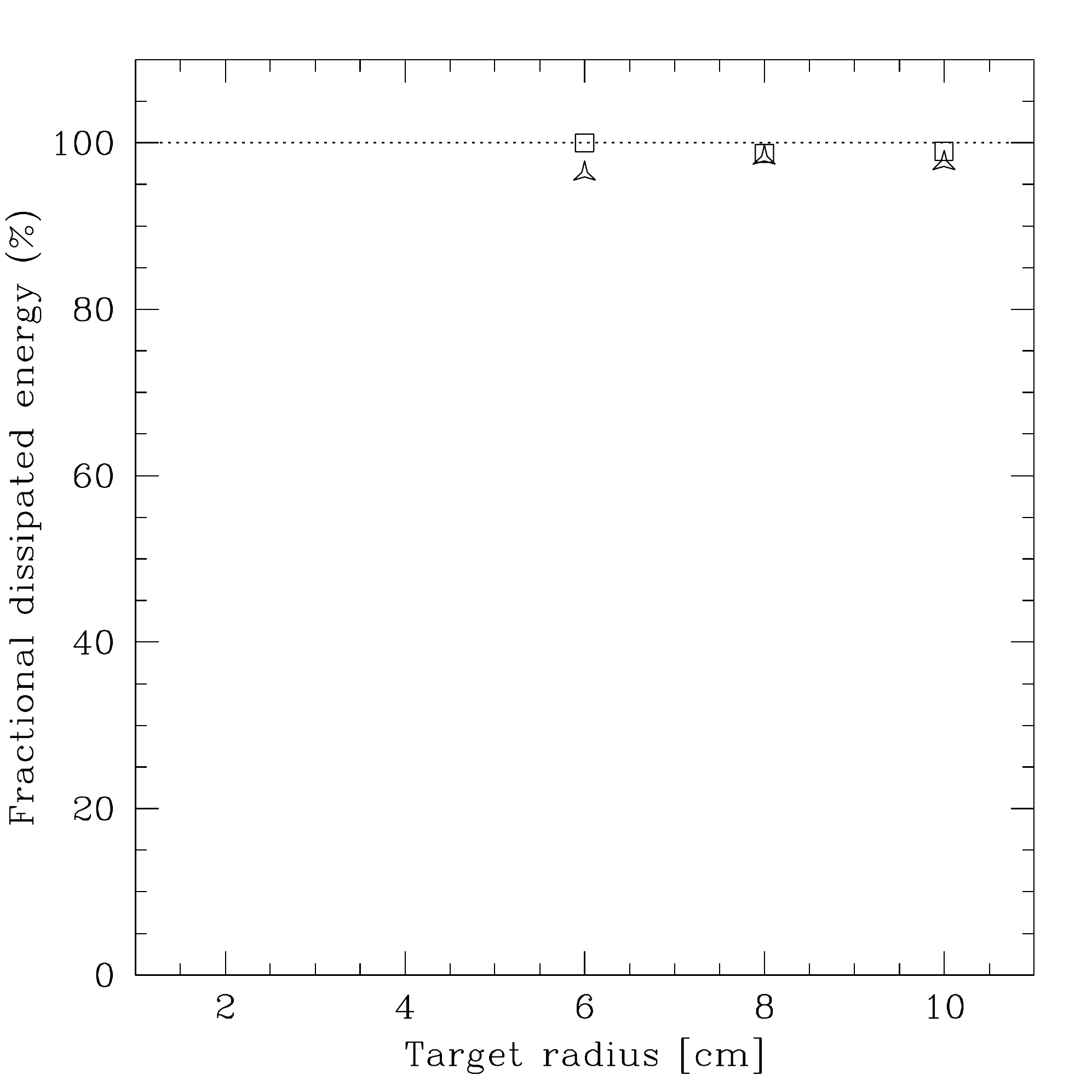}
  \caption{Graph showing the energy dissipated as a percentage of the maximum available energy in the collision against the target radius, for each of the simulations carried out that are closest to the fragmentation velocity that involve growth (squares) and fragmentation (triangles).  The dotted line marks where perfect sticking can occur (Equation~\ref{eq:Ediss}) i.e. the maximum amount of energy that can be dissipated.  The simulation data will always be at or below this maximum line.  The closer the data point is to 100\%, the closer it is to the threshold velocity.  The projectile radius is $\Rp = 6$~cm and the aggregate initial filling factors are 35\%.}
\label{fig:ediss_tsize_boundary_analytical}
\end{figure}

Table~\ref{tab:sim_mass_ratio} (upper panel) and Figure~\ref{fig:vth_tsize} summarise how the fragmentation velocities change with target size.  Since we have just three data points in Figure~\ref{fig:vth_tsize}, it is not possible to determine the exact functional form but it is clear that as the target mass increases, the fragmentation velocity also increases.  As mentioned in Section~\ref{sec:mproj} this is because a collision between more extreme mass ratio aggregates allows energy to be absorbed in elastic loading and then dissipated more efficiently via plastic deformation, resulting in a higher threshold velocity for fragmentation.  Figure~\ref{fig:ediss_tsize_boundary_analytical} shows that the fragmentation velocities obtained are indeed close to the perfect sticking case.

\subsubsection{Constant size/mass ratio}

\begin{figure}
\centering
  \includegraphics[width=1.0\columnwidth]{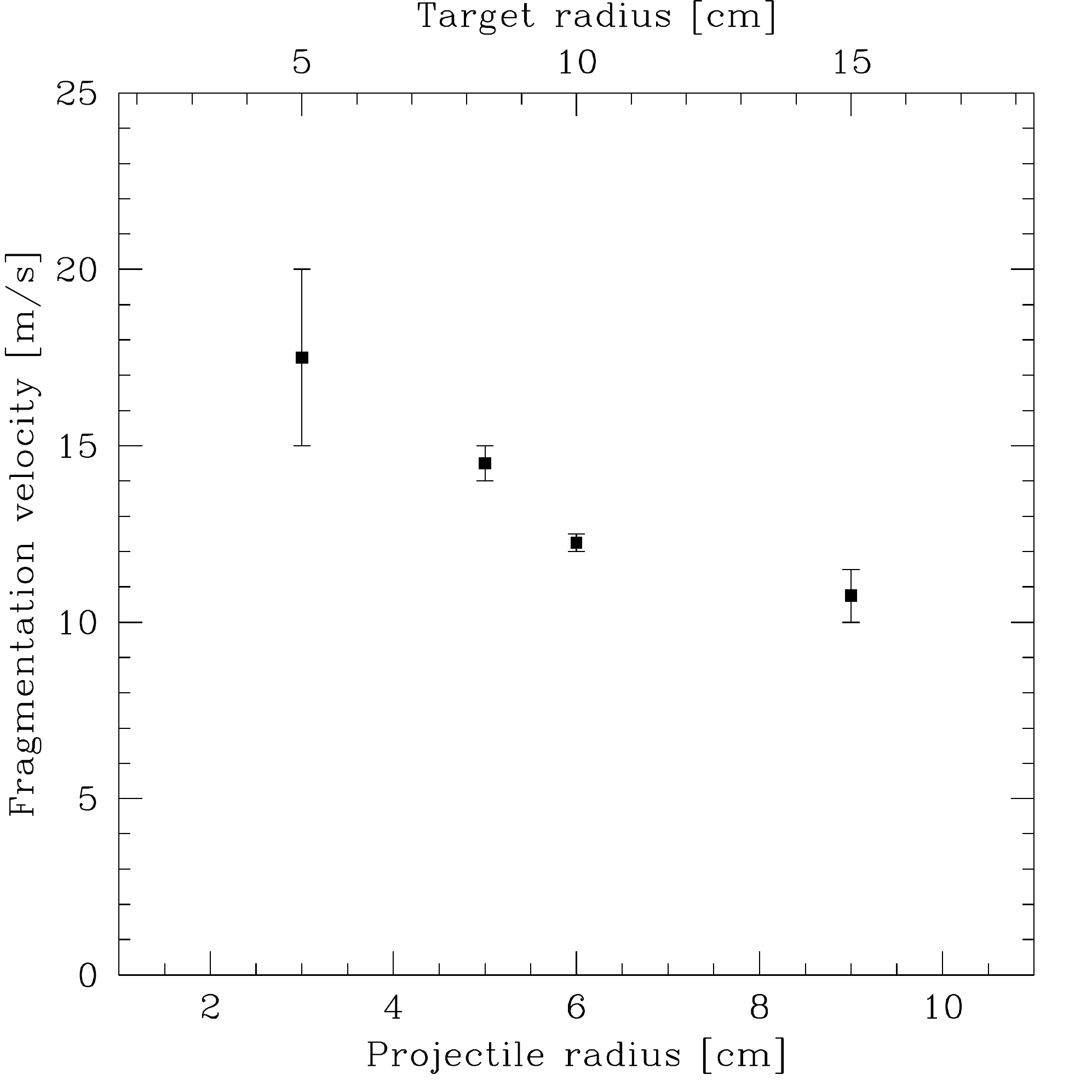}
  \caption{Graph showing the fragmentation velocity against projectile and target radii, changed in proportion to maintain a fixed mass ratio.  The error bars show the upper and lower limits around the fragmentation velocities, obtained from the simulation results.  The threshold velocity for fragmentation decreases as the projectile and target sizes are increased.  The simulations are carried out with filling factor, $\phi = 35\%$.}
\label{fig:vth_mass_ratio}
\end{figure}

\begin{figure}
\centering
  \includegraphics[width=1.0\columnwidth]{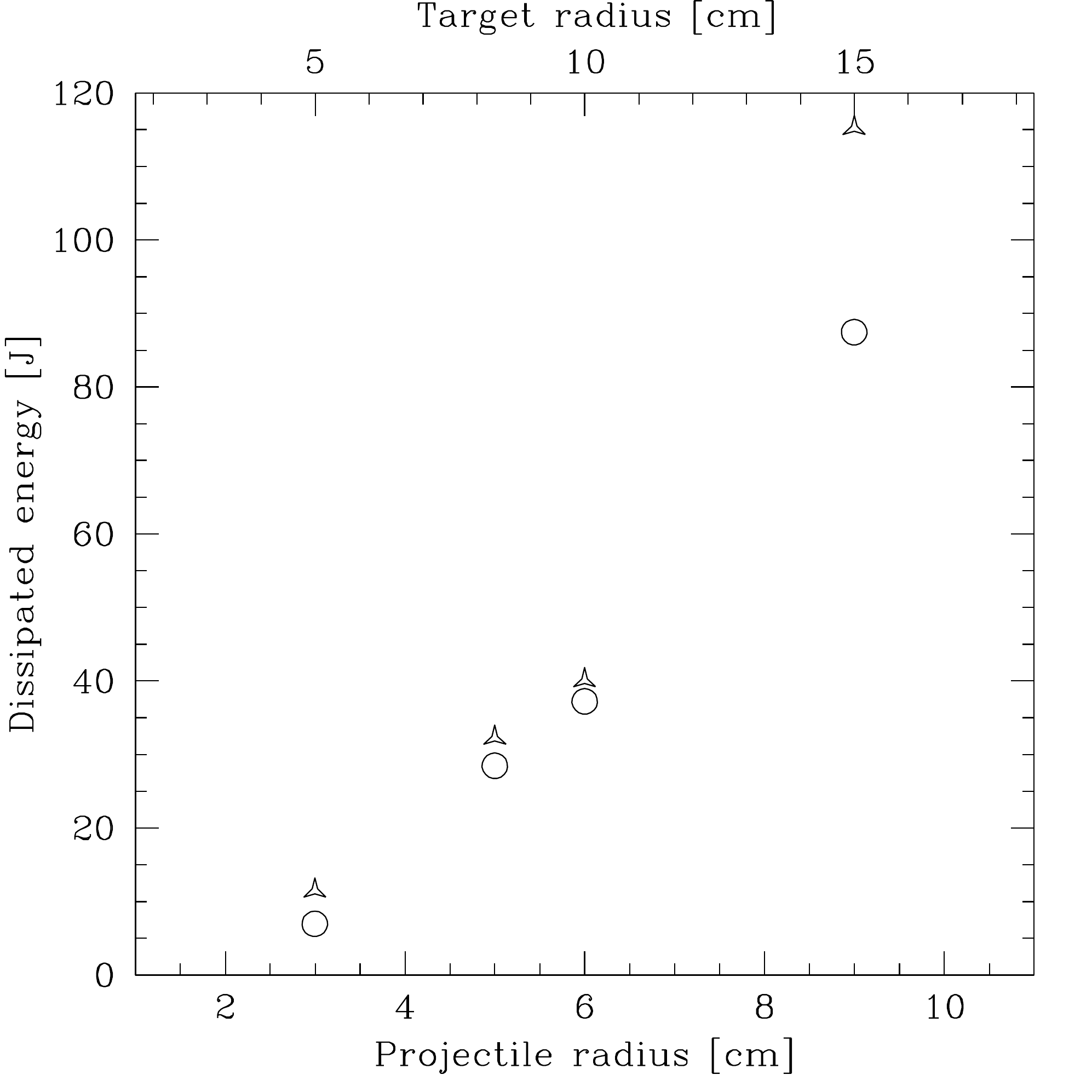}
  \caption{Graph showing the absolute energy dissipated against projectile and target radii, changed in proportion to maintain a fixed mass ratio, for each of the simulations carried out that are closest to the fragmentation velocity that involve growth (circles) and fragmentation (triangles).  The initial energy varies in each of these simulations since the projectile masses are changed.  The amount of energy that can be dissipated increases as the projectile and target masses are increased.  The simulations are carried out with filling factor, $\phi = 35\%$.}
\label{fig:abs_ediss_boundary}
\end{figure}

Table~\ref{tab:sim_mass_ratio} (lower panel) and Figure~\ref{fig:vth_mass_ratio} show the effect of changing the size of the projectile and target in proportion (such that the mass ratio remains constant) on the fragmentation velocities.  Increasing the aggregate masses in proportion increases the energy available in the collision for any one collision velocity (Equation~\ref{eq:Ediss}).  It can be clearly seen that as the aggregate sizes increase the threshold velocity for fragmentation decreases.  However, while the projectile mass increases by a factor of 27 between the smallest and largest set of sizes considered, the fragmentation velocity only decreases by a factor of $\approx 1.6$.  Therefore, for perfect sticking the energy required to be dissipated in a collision increases as the projectile and target mass increase, suggesting that as aggregates become larger, the likelihood of sticking is smaller.  Figure~\ref{fig:abs_ediss_boundary} shows that the absolute energy dissipated decreases by a factor of $\approx 10$ when the projectile and target masses are decreased between the largest and smallest set of sizes considered.  Therefore, as an object grows larger, it becomes more efficient at dissipating energy (as shown in Section~\ref{sec:mtarget}).  However, since there is more energy to dissipate due to the increased mass, the fragmentation velocity decreases (equation~\ref{eq:Ediss}), the end result being that the larger aggregates cannot withstand high collision velocities (as shown in Section~\ref{sec:mproj}).  Thus, it is clear that both the dissipated energy and the fragmentation velocity vary, and whether the two aggregates will collide and grow will depend on both the mass and the aggregates' ability to dissipate energy.

Note that if the mass ratio between the colliding objects remains constant, the specific energy is the same for any one collision velocity.  Therefore, one might expect that the fragmentation velocity should remain the same.  However, we point out that the non-linear strength relations do mean that a dependence on the velocity is possible, as seen in Figure~\ref{fig:vth_mass_ratio}.

\subsubsection{Comparison with the catastrophic disruption threshold}

\begin{figure}
\centering
  \includegraphics[width=1.0\columnwidth]{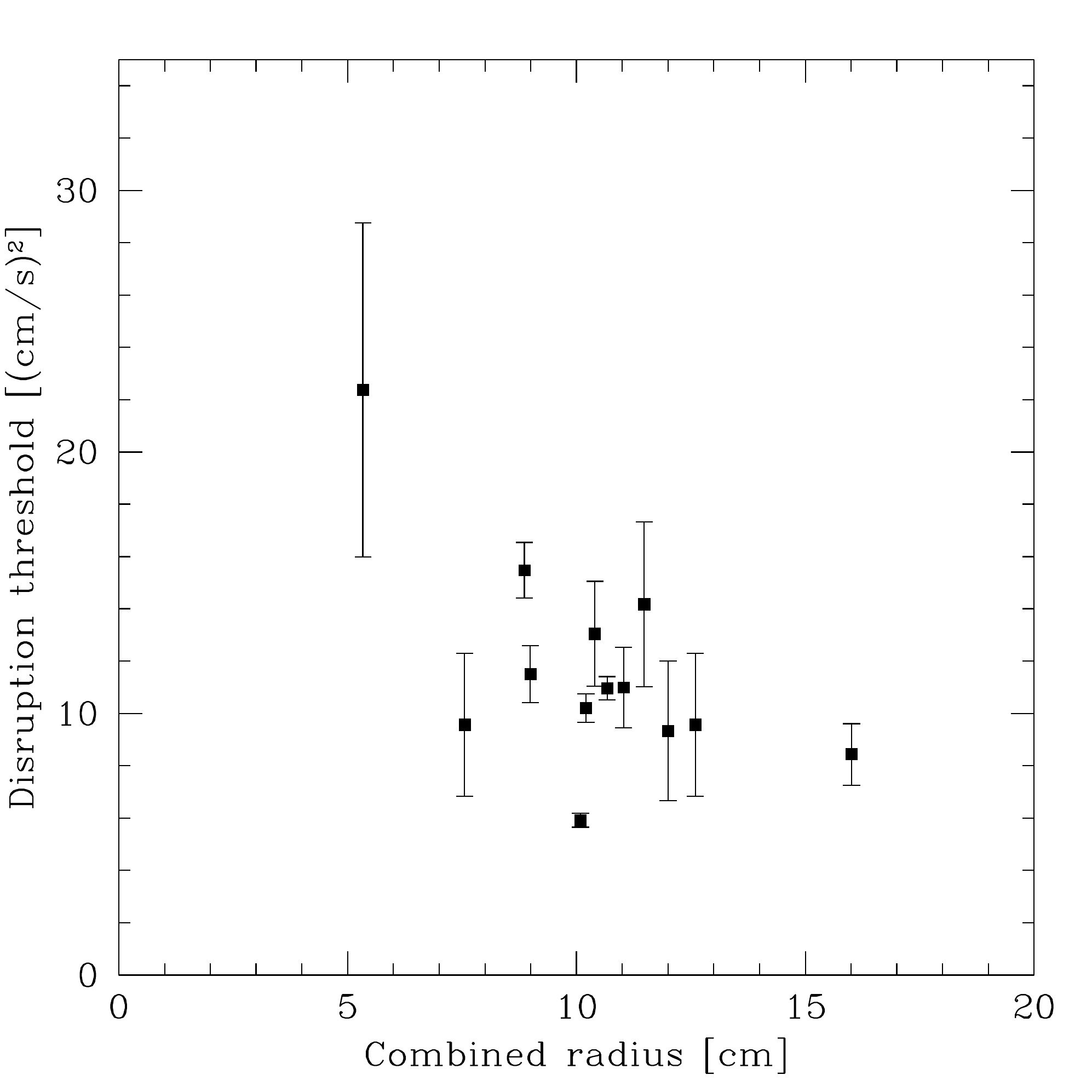}
  \caption{Graph showing a proxy for the catastrophic disruption threshold given by equation~\ref{eq:Qproxy} against the combined radius for the simulations performed in this paper with initial filling factors of $\phi = 35\%$.  As the combined radius increases, the disruption threshold decreases.}
\label{fig:Qstar_Rc}
\end{figure}

The catastrophic disruption threshold is usually defined as the kinetic energy of the projectile divided by the target mass required to break the target such that only half of its initial mass remains.  However, since this definition does not account for the mass of the projectile, \cite{Stewart_Leinhardt_Qstar_CofM} proposed an alternative disruption threshold which is the reduced mass kinetic energy divided by the total system mass (or equivalently the specific energy in the centre of mass frame) that is required to ensure that the largest fragment is only half of the total mass.  This therefore accounts for the case where the projectile is of a similar mass to the target (as in our simulations).  The reduced mass kinetic energy (or equivalently, the initial impact energy in the centre of mass frame) is given by equation~\ref{eq:Ediss}.  Though our fragmentation velocity is defined to be the lowest velocity at which the largest fragment is smaller than the initial target mass, we can use the threshold values as a proxy to determine the catastrophic disruption values as follows:

\begin{equation}
Q_{\rm proxy, CofM} = \frac{1}{2} \frac{\mu v_{\rm th}^2}{\mproj + \mt}.
\label{eq:Qproxy}
\end{equation}
where $\mu = \mproj\mt/(\mproj + \mt)$ is the reduced mass of the system.  Figure~\ref{fig:Qstar_Rc} shows the catastrophic disruption threshold given by equation~\ref{eq:Qproxy} against the combined radius (defined to be $R_{\rm c} = (\Rt^3 + \Rp^3)^{1/3}$, as done so by \cite{Stewart_Leinhardt_Qstar_CofM}), for all the simultions carried out in this paper using a filling factor of $\phi = 35\%$.  One can see that the general trend is for a decrease in the catastrophic threshold with combined radius.  This is consistent with the trend observed by \cite{Stewart_Leinhardt_Qstar_CofM} from their numerical modelling (see their Figure 2).  However, due to our large error bars and lack of spread of data in the combined radius in Figure~\ref{fig:Qstar_Rc}, we are unable to constrain the fit exactly.

\subsection{Aggregate porosity}
\label{sec:porosity}

\begin{figure}
\centering
\includegraphics[width=1.0\columnwidth]{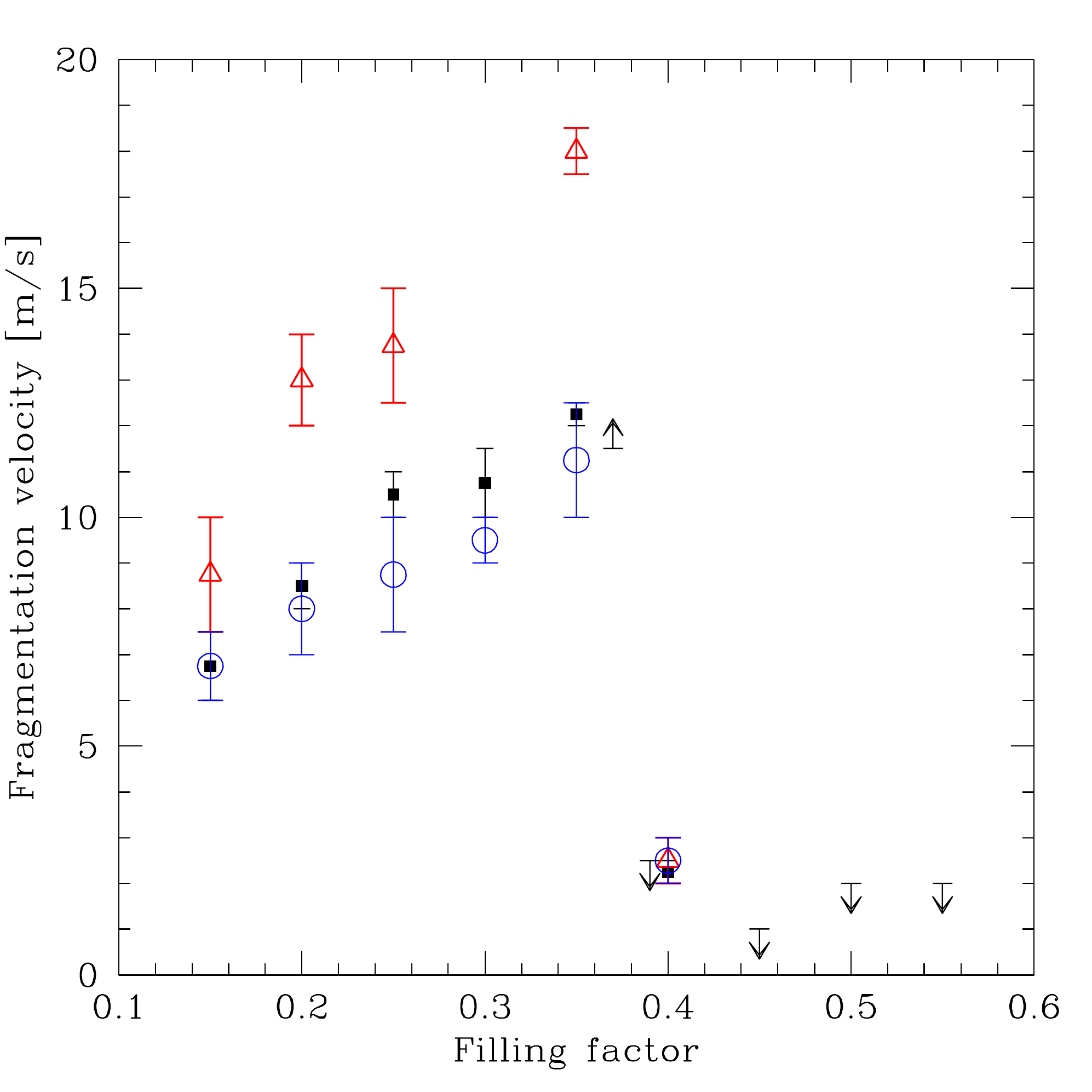}
  \caption{Graph showing the fragmentation velocity against filling factor for simulations carried out with $\Rt = 10$~cm and $\Rp = 6$~cm (black squares), $\Rp = 4$~cm (red triangles) and $\Rp = 8$~cm (blue circles).  The error bars show the upper and lower limits around the threshold velocities, obtained from the simulation results.  Where only the upper or lower limit exists, it is shown using a downwards or upwards arrow, respectively.  The fragmentation velocity increases as the filling factor increases until a filling factor of $\approx 37\%$ where a sharp drop in the fragmentation velocity occurs for all three sizes considered.}
\label{fig:vth_porosity}
\end{figure}

Table~\ref{tab:sim_porosity} and Figure~\ref{fig:vth_porosity} summarise the effects that the aggregate porosity has on the fragmentation velocity.  As the aggregate filling factor increases, the threshold velocity above which fragmentation occurs also increases.  Beyond a particular filling factor ($\approx 37\%$) the threshold velocity for fragmentation rapidly drops.

In general, as the filling factor increases the magnitude of the strength quantities, i.e. tensile, compressive and shear strengths, also increase (equations~\ref{eq:tensile-strength}, \ref{eq:compressive-strength} and~\ref{eq:shear-strength}, respectively; see also Figure~\ref{fig:strengths}).  We can consider this in the context of the results for filling factors, $\phi \lesssim 37\%$.  The compressive, tensile and shear strengths play very different and opposing roles at different filling factors.  At low filling factors (i.e. low strengths) the elastic regime is small (see Figure~\ref{fig:strengths}).  Therefore, plastic deformation sets in for small stresses (i.e. small impact energies).  Plastic compression is an effective energy dissipation mechanism, especially for highly porous aggregates.  When material is being compacted at low filling factors, the compressive strengths result in the impact energy being dissipated mostly by plastic deformation.  This results in a material with a higher filling factor and therefore since the aggregate moves into a regime where the elastic regime is larger, it becomes more stable.  On the other hand, the tensile and shear stresses cause the material to be easily deformed and destroyed.  At large filling factors (i.e. large strengths), the plastic regime only sets in for larger impact energies and thus a larger elastic regime exists in comparison to the low filling factor case.  The larger compressive strengths mean that the impact energy is not dissipated as
efficiently as for low filling factors but is instead stored in
the elastic loading of the aggregates, which can potentially contribute to the destruction of the aggregates.  On the other hand, the higher plasticity limit makes the objects more stable against tensile and shear disruption, making the aggregates less fragile.

\begin{table*}
\centering
\begin{tabular}{llll}
    \hline
    Strength & Low filling factors & High filling factors\\
    \hline
    \hline
   Compressive & Easily dissipate energy (good for growth) & Cannot dissipate energy easily (bad for growth)\\
   Tensile & Allows aggregate weakening \& destruction (bad for growth)& Can resist tensile disruption (good for growth)\\
    Shear & Allows aggregate weakening \& destruction (bad for growth) & Can resist shear disruption (good for growth)\\    
    \hline
 \end{tabular}
 \caption{Table summarising the effects that the compressive, shear and tensile strengths have on the aggregates at low and high filling factors.}
 \label{tab:stability}
\end{table*} 

\begin{figure}
\centering
  \includegraphics[width=1.0\columnwidth]{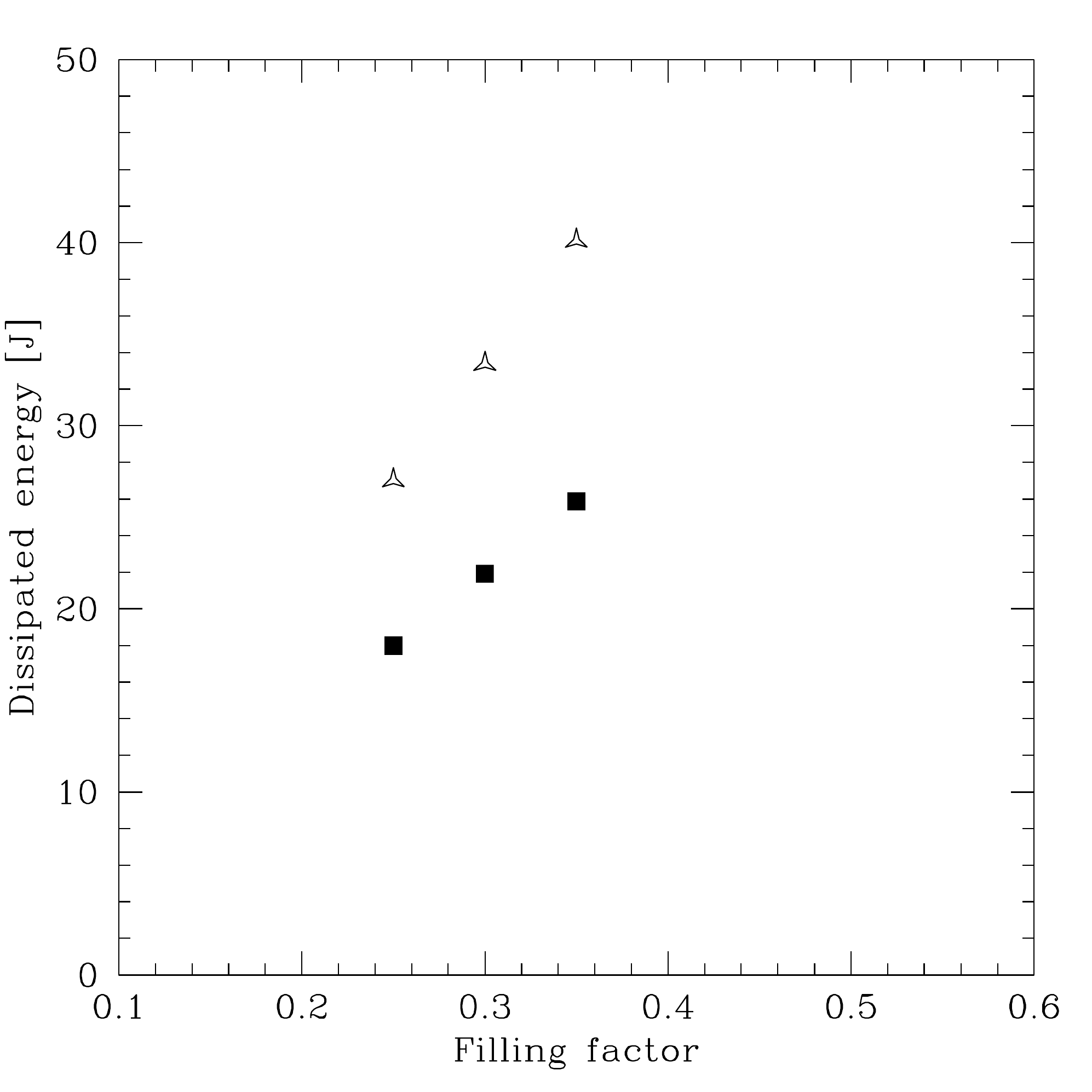}
 \caption{Graph showing the absolute energy dissipated against filling factor for a set of low filling factor simulations carried out with collision velocities of $10$~m/s (solid squares) and $12$~m/s (open triangles).  As the filling factor increases the aggregate's ability to dissipate energy also increases due to its larger mass.  The simulations carried out at $10$~m/s are growth simulations while those carried out at $12$~m/s are destructive simulations.  The target and projectile have radii $\Rt = 10$~cm and $\Rp = 6$~cm, respectively.}
\label{fig:Eabs_constVc}
\end{figure}

In summary, at low filling factors ($\lesssim 37\%$), the compressive strength acts to enhance growth of the aggregates while the tensile and shear strengths act to break them apart, while the converse is true in the high filling factor case, resulting in competing processes (Table~\ref{tab:stability}).  Overall the aggregates gain a higher stability as the filling factor is increased and the fragmentation velocity increases further for higher filling factors.  Due to the stronger structure, more kinetic energy can be stored in elastic loading of the aggregate (note that the higher filling factor also means that the aggregate mass is larger so the total initial energy is also higher).  To illustrate this, we measure the dissipated energies in three simulations carried out with a collision velocity of $10$~m/s (in which growth occurs) as well as $12.5$~m/s (in which negative growth occurs).  Figure~\ref{fig:Eabs_constVc} shows that as the filling factor increases the dissipated energy also increases.  Therefore with an increased filling factor, an aggregate's ability to dissipate energy also increases due to its increased mass.

\begin{figure}
\centering
  \includegraphics[width=1.0\columnwidth]{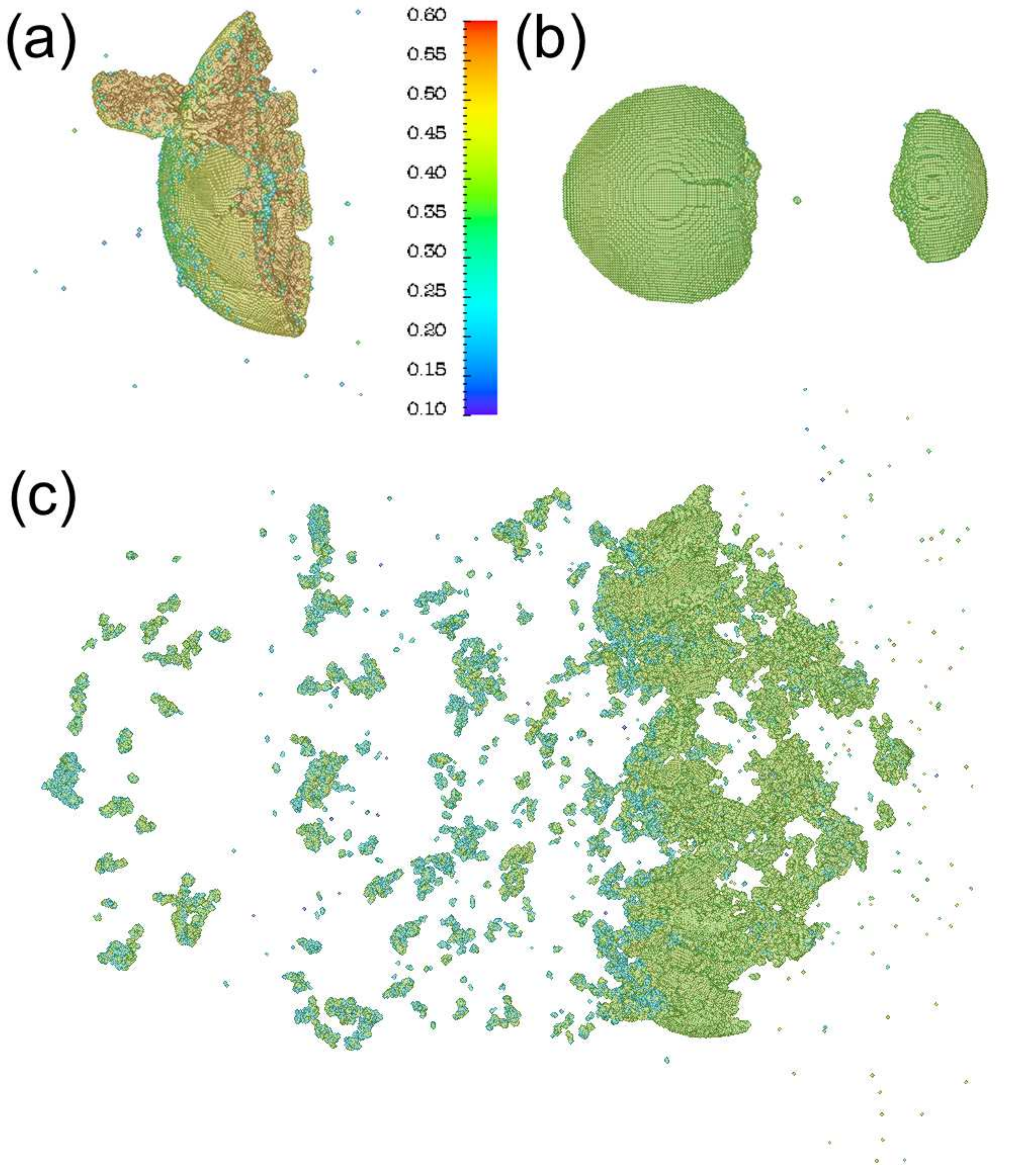}
  \caption{Filling factor rendered images illustrating the result of collisions close to the fragmentation velocity drop at $\phi \approx 37\%$.  (a) For $\phi = 37\%$ and $\vc = 11.5$~m/s the projectile sticks to the target and a smaller chunk breaks off.  (b) For $\phi = 40\%$ and $\vc = 2.0$~m/s the aggregates rebound with some mass transfer onto the target and smaller parts chipping off. (c) However, for the same filling factor at a higher collision velocity ($\vc = 5$~m/s) the aggregates fragment.  The target and projectile radii are $\Rt = 10$~cm and $\Rp = 6$~cm, respectively.}
\label{fig:porosityimage}
\end{figure}

At a filling factor of $\phi \approx 37\%$ a sudden drop in the fragmentation velocity occurs.  The effects of this sharp drop are also evident in Figure~\ref{fig:porosityimage}: at $\phi \approx 37\%$ and a high velocity of $\vc = 11.5$~m/s the growing aggregate is still intact after the collision and most of the projectile mass sticks to it (Figure~\ref{fig:porosityimage}~a).  For $\phi = 40\%$ the aggregates essentially rebound off each other at a collision velocity of $\vc = 2$~m/s (Figure~\ref{fig:porosityimage}~b) but for a higher collision velocity of $\vc = 5$~m/s the aggregates fragment (Figure~\ref{fig:porosityimage}~c). This indicates a significantly enlarged loss regime compared to the situation for $\phi = 37\%$.

At first this drop is surprising.  It is caused by a complex interplay between the elastic and plastic properties of the aggregates which makes it difficult to capture this feature quantitatively, though results on the bouncing behaviour of dust aggregates by \citet{Four_population} and \citet{Geretshauser.2012} can provide some insight into the processes occurring here.  When increasing the collision velocity these authors find three types of threshold velocity transitions related to the growth conditions:

(i) a gain-loss threshold (i.e. a transition from positive to negative growth) where at low velocities the largest aggregate grows while at higher velocities the aggregates fragment.

(ii) a neutral-gain transition: a velocity threshold below which the aggregates rebound and above which the largest aggregate grows.  Note that in this paper we do not explore this region of the parameter space in detail but three simulations that are carried out in this regime (see bold entries in Table~\ref{tab:sim_mass}) are consistent with a neutral-gain transition.

(iii) a neutral-loss transition: a threshold velocity below which the aggregates bounce and above which the aggregates fragment.  This particularly occurs at high filling factors.  Note that we see evidence of this in our high filling factor simulations (see bold entries in Table~\ref{tab:sim_porosity}).

All three transitions are also seen in laboratory experiments \citep{Guttler.2010}.  Since we focus on a velocity threshold that separates positive from negative growth, the data points for $\phi < 37\%$ in Figure~\ref{fig:vth_porosity} indicate a gain-loss threshold.  However, the results by \citet{Four_population} and \citet{Geretshauser.2012} indicate that there is also a neutral-gain transition in this filling factor regime. For $\phi = 35\%$ they find a threshold of $\approx 1$~m/s for this transition, which extends to the region in which $\phi > 37\%$.  For $\phi > 37\%$ the graph represents a neutral-loss transition (at $\approx 1$~m/s) while the gain region is very narrow or completely vanishes.

We identify the following aspects that may contribute to this sudden disappearance of the gain region:

(1) As the filling factor increases, the compressive strength $\Sigma(\phi)$ increases resulting in an increasing fraction of the initial kinetic energy being stored in elastic loading of the aggregates.  Consequently, this elastic energy is available for the separation of the aggregates in a bouncing collision.  Moreover, when the filling factor reaches $\phi=0.58$, the compressive strength is no longer limited (see Figure~\ref{fig:strengths}). Hence no more energy can be dissipated by plastic compression in this filling factor regime.

(2) The increase in filling factor also causes the bulk modulus of the aggregates to increase (Equation~\ref{eq:bulk-modulus}) resulting in stiffer aggregates.  This causes the contact area between the aggregates to decrease and less energy is necessary to separate the aggregates in a bouncing collision.

(3) Since there will always be partial sticking of aggregates, a clean neutral bouncing is unrealistic. Instead some mass transfer or even the rip out of larger chunks can be expected (see Figure~\ref{fig:porosityimage}~b).

(4) Due to the increase in $\Sigma(\phi)$ and $K(\phi)$ density waves of increasing amplitude propagate across the aggregate and lead to a local rarefaction of the material and consequential local reduction of shear and tensile strength. These density waves finally rip the aggregate apart even at low collision velocities.

As the filling factor is increased, aspects (1) and (2) decrease the amount of sticking of the projectile to the target and increase the amount of bouncing of the aggregates.  A change of behaviour occurs when the filling factor of the aggregate reaches $\phi=58\%
(=\phi_2$; see Equation~\ref{eq:compressive-strength}), which is the
porosity limit for {$\SiOtwo$}. At this point the compressive strength
becomes infinite. This conforms to a phase transition of the material
from porous to completely compact. A porous material can dissipate
energy by compaction, while compact objects cannot.  By the loss of this
channel of dissipation, the material is dramatically more vulnerable
against fragmentation.  Although the resistance against tensile and
shear disruption also increases, the lack of energy dissipation by
compaction dominates the fate of the aggregates during collisions.  In addition, aspect (4) means that fragmentation of the aggregate increases with increasing filling factor. The low velocity collisions at $\phi \gtrsim 37\%$ in Figure~\ref{fig:vth_porosity} show a behaviour where the aggregates mainly bounce off each other but partly also fragment or some mass transfer (aspect 3) occurs (also see Table~\ref{tab:sim_porosity}).  Therefore a region of small positive or negative growth may formally be identified, though this is the boundary region of a neutral regime.  Since sticking (which results in positive growth) and bouncing (potentially with some fragmentation or mass transfer, resulting in approximately neutral growth) are both distinctly different physical processes, a sudden change in the fragmentation velocity at the boundary between both regimes is not unreasonable.  We note that a collision model by \cite{Guttler.2010} based on experimental data features only the categories ``porous'' and ``compact'' aggregates and uses a filling factor of $\phi = 40\%$ as a separation criterion which is close to the filling factor of $\phi \approx 37\%$ where the steep drop in the fragmentation velocity occurs.  However, this criterion is selected quite arbitrarily by splitting the possible filling factor regime between 0.15 and 0.65 into two halves.  The steep drop that we observe now provides a physical basis for this choice. Other experimental investigations find that the maximum filling factor that can be reached for porous aggregates under protoplanetary disc conditions is $\phi \approx 33\%$ \citep{Teiser.2011}. However, this value is obtained for polydisperse $\rm SiO_2$ dust which is less compressible than the monodisperse dust used to calibrate our SPH code.  This value can therefore be expected to be higher for monodisperse dust, as we observe.

To summarise, the drop in our fragmentation velocity curve occurs where experimental data also indicates a significant change in the collision behaviour of dust aggregates.  This drop originates from the phase transition of the
material at the theoretical porosity limit of $\SiOtwo$ at filling
factor $\phi = 58\%$. As soon as the possibility to dissipate energy by
plastic compaction ceases, the aggregates are far more easily
disrupted.  For the simulations with $\phi>37\%$, parts of the aggregate
may already pass this limit during the initial impact phase for impact
velocities as low as $\approx$~$1-2$~$\mathrm{m/s}$, while for lower
initial filling factors this limit is not reached during the initial
impact.  To test this, we have performed additional simulations where we set the
compressive strength, $\Sigma(\phi)$, to infinity for all filling factors,
$\phi$. The results of these simulations resemble those
with $\phi>37\%$ (whereby energy cannot be dissipated and instead disruption occurs), indicating the dominant role of plastic compression
for energy dissipation.

\subsubsection{Porosity of fragments following a destructive collision}

\begin{figure*}
\centering
  \includegraphics[width=1.0\columnwidth]{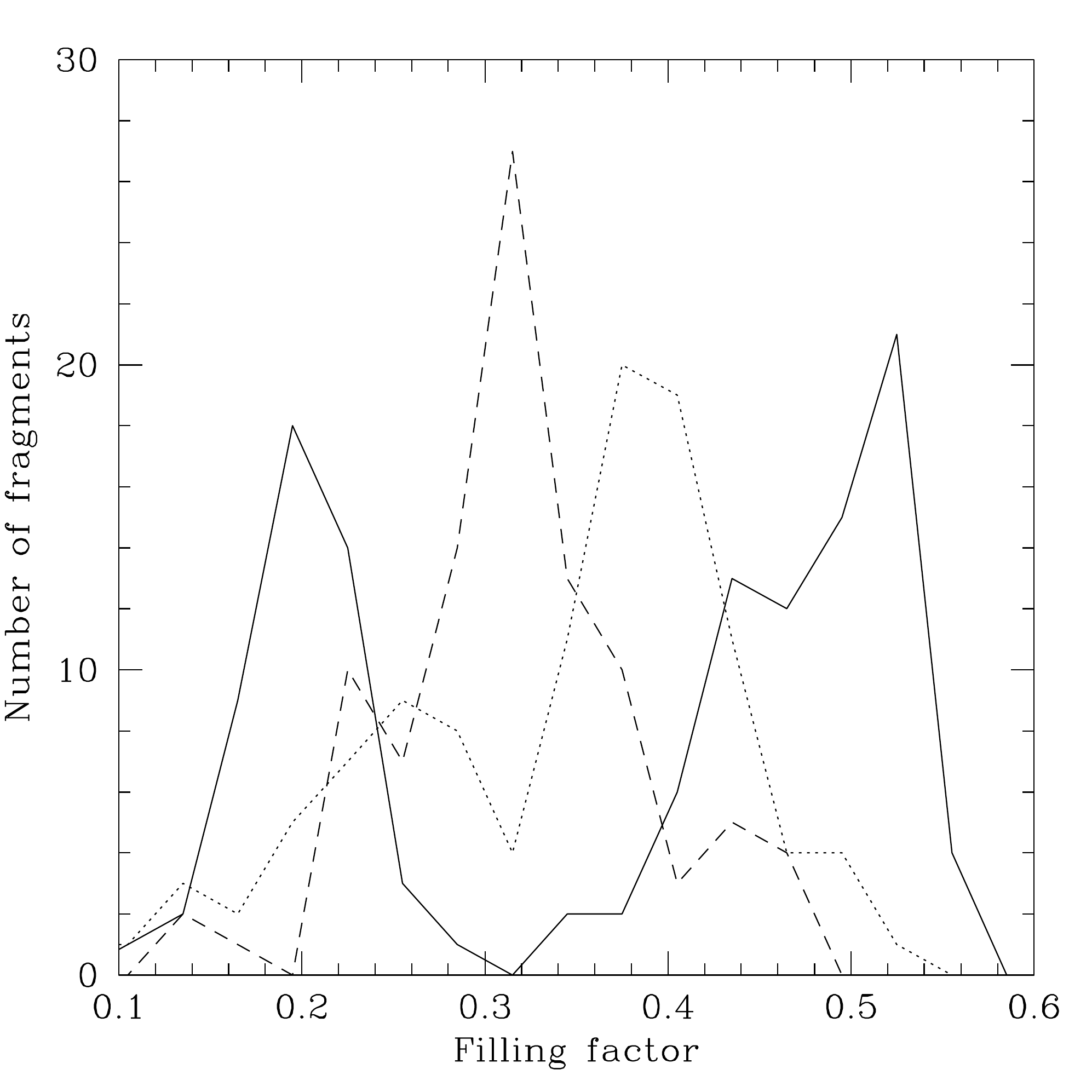} \includegraphics[width=1.0\columnwidth]{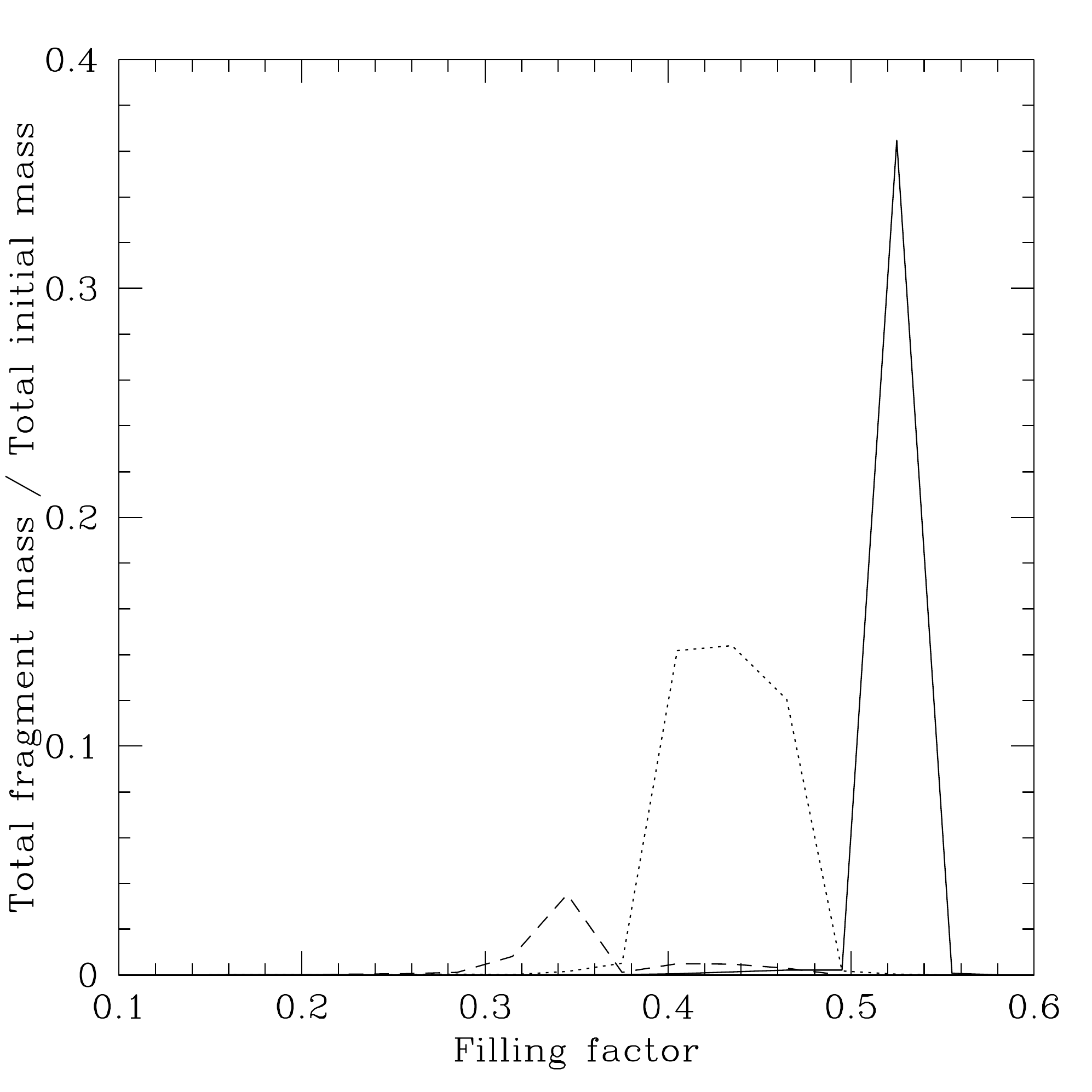}
  \caption{Graph showing the number (left panel) and mass (right panel) distribution of the final filling factor of fragments formed in a destructive collision with initial filling factors $\phi = 15\%$ (dashed line), $25\%$ (dotted line) and $35\%$ (solid line) with velocities that are just on the fragmenting side of the threshold.  For the low filling factor most of the fragments are compressed before breaking off, while for the higher filling factors a bimodal distribution of the number of fragments occurs.  The fragments with a lower filling factor are likely to be the result of being chipped off in the collision.  However, most of the mass is in the higher filling factor bins.  The initial target and projectile radii are $\Rt = 10$~cm and $\Rp = 6$~cm, respectively.}
\label{fig:dens_dis_lowff}
\end{figure*}

\begin{figure*}
\centering  \includegraphics[width=1.0\columnwidth]{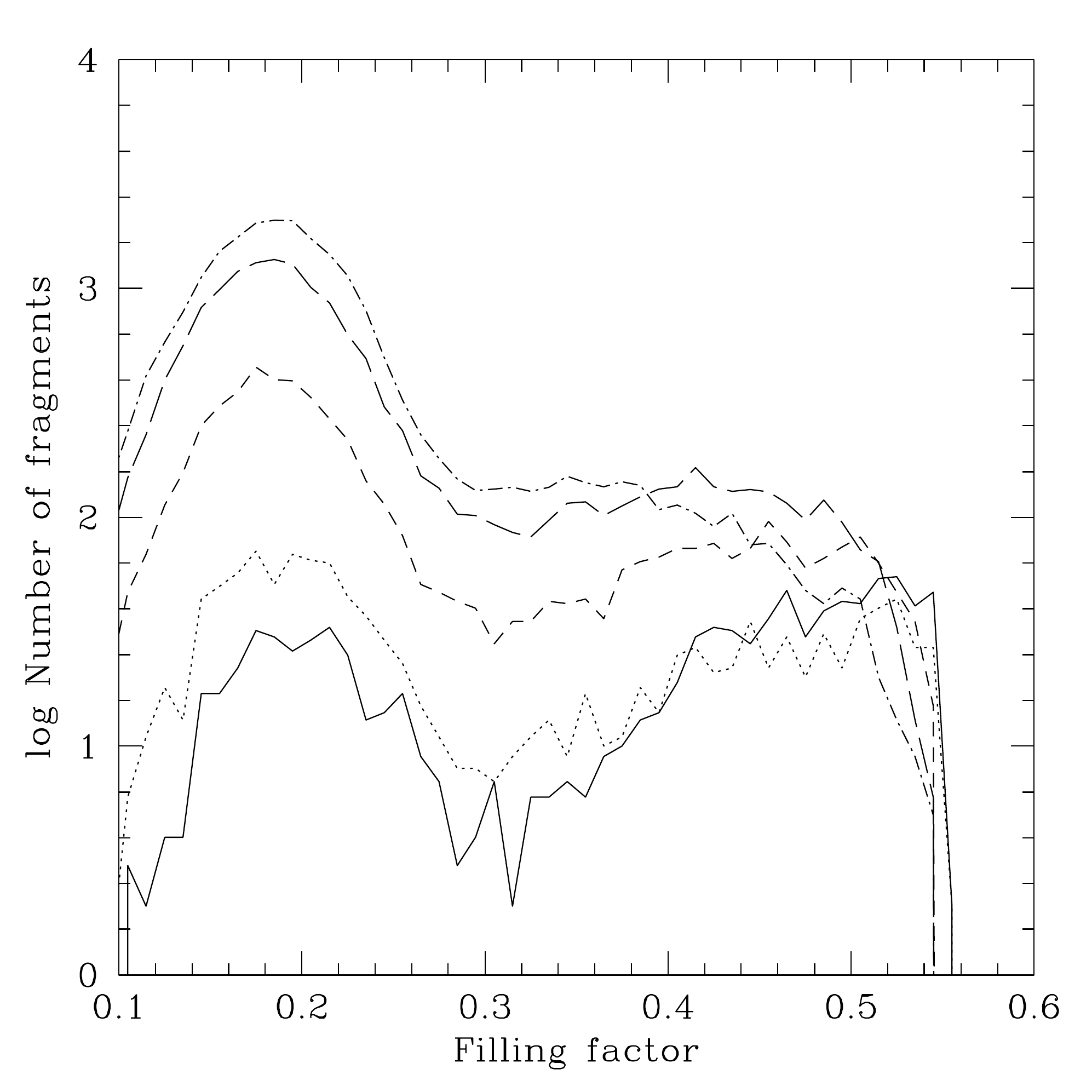} \includegraphics[width=1.0\columnwidth]{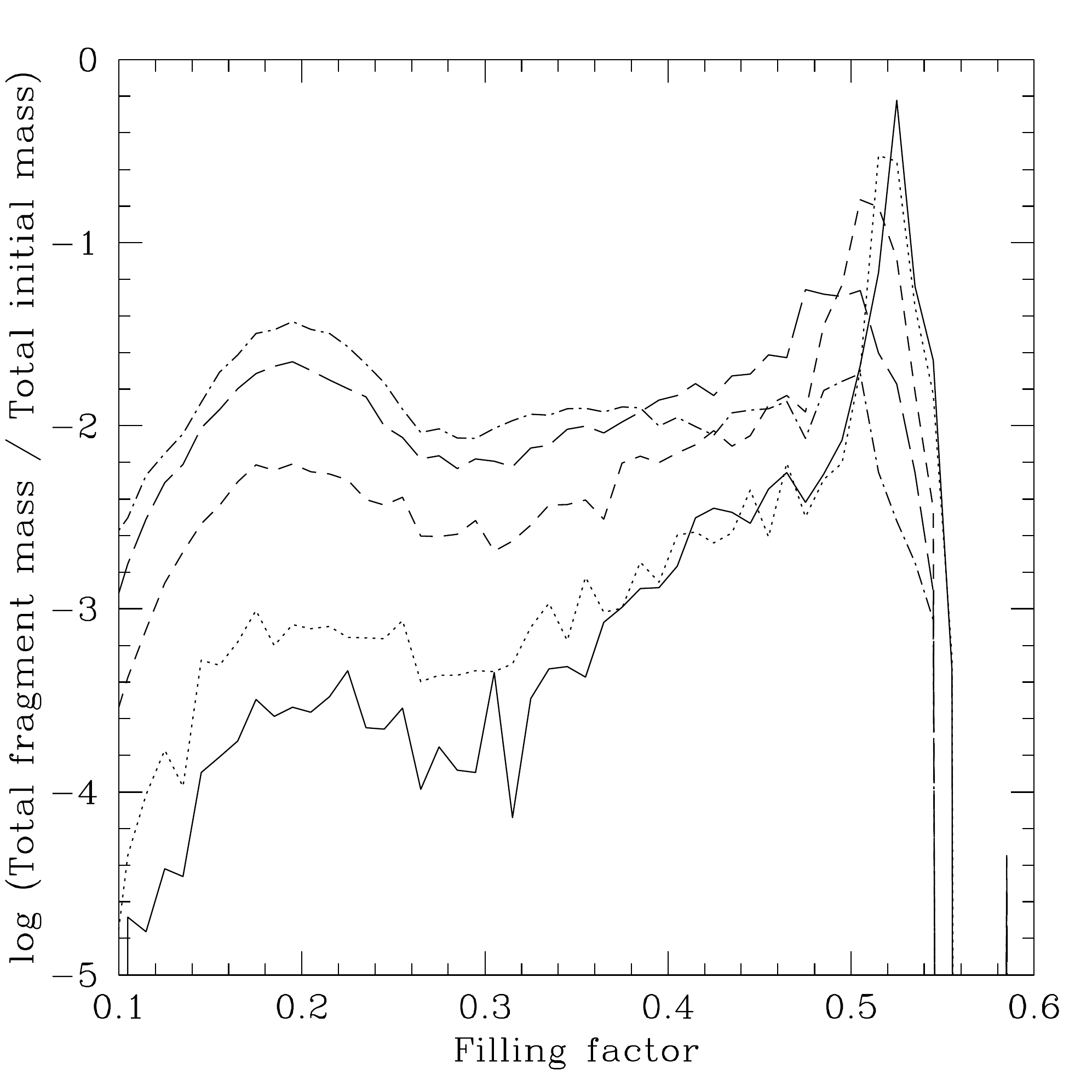}
  \caption{Graph showing the number (left panel) and mass (right panel) distribution of the final filling factors of fragments formed in a destructive collision with collision velocities $\vc = 17.5$ (solid line), 20.0 (dotted line), 22.5 (short dashed line), 25.0 (long dashed line) and 27.5 (dot-dashed line) m/s.  As the collision velocity is increased the number of fragments, as well as the proportion of mass, in the lower filling factor bins increases.  The proportion of mass in the higher filling factor bins also decreases.  The initial target and projectile radii are $\Rt = 10$~cm and $\Rp = 6$~cm, respectively, and the initial filling factor is $\phi = 35\%$.}
\label{fig:dens_0.35_high_vcoll}
\end{figure*}

\begin{figure*}
\centering  \includegraphics[width=1.0\columnwidth]{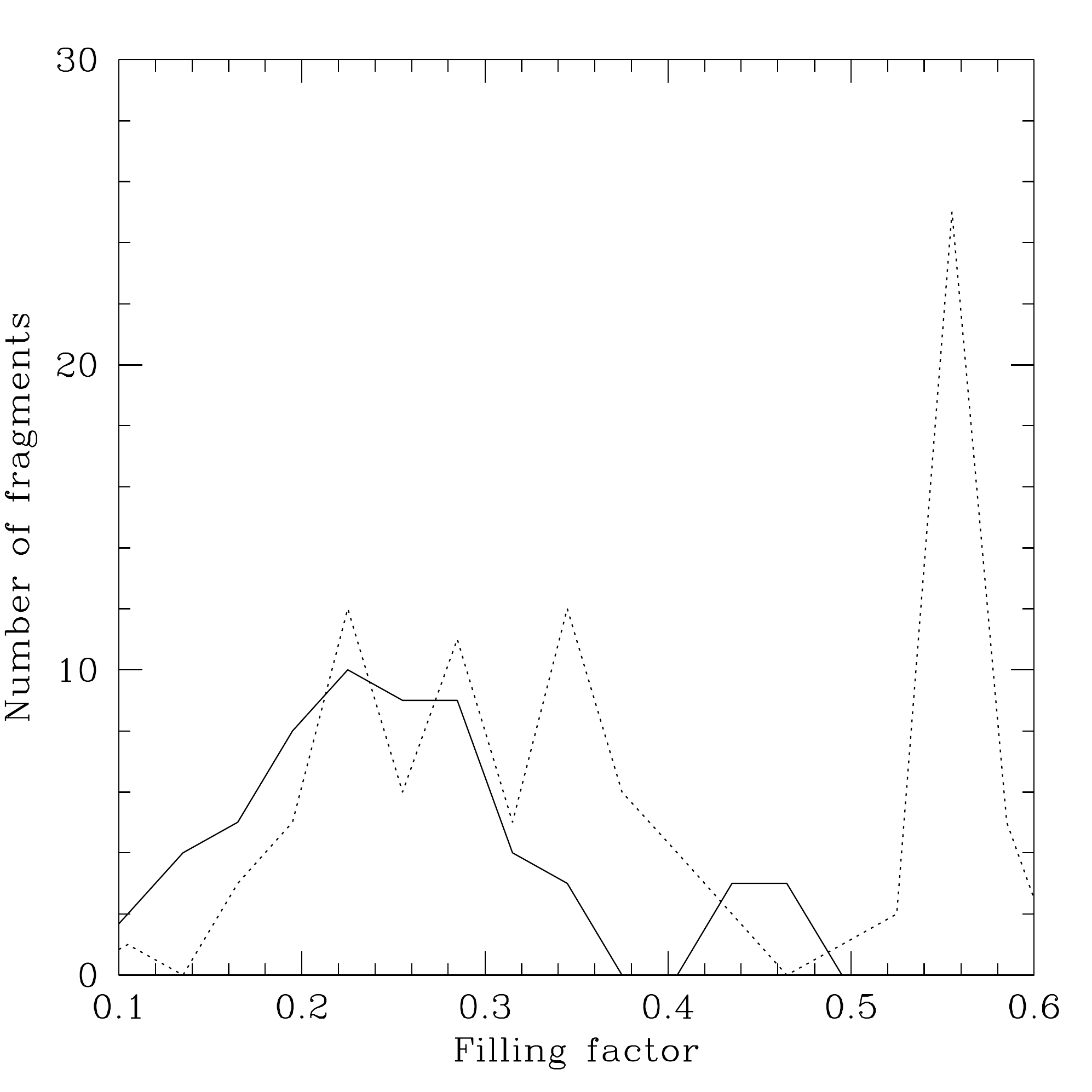} \includegraphics[width=1.0\columnwidth]{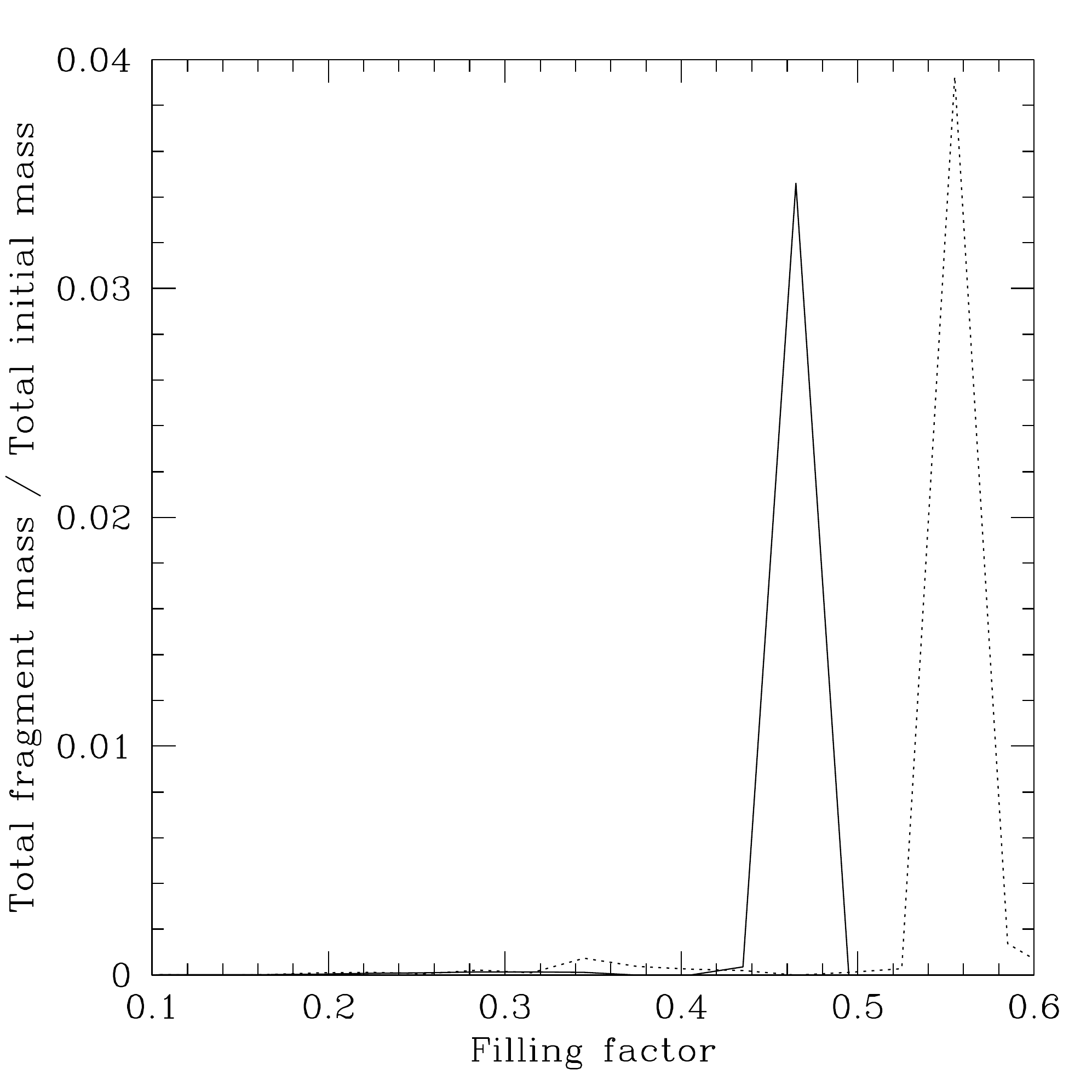}
  \caption{Graph showing the number (left panel) and mass (right panel) distribution of the final filling factors of fragments formed in destructive collisions with $\vc = 2.5$~m/s and initial filling factors $\phi = 45\%$ (solid line) and $55\%$ (dotted line).  While a decrease in filling factor does occur for some of the fragments, many maintain the initial filling factor.  However, most of the mass is in the higher filling factor bins.  The initial target and projectile radii are $\Rt = 10$~cm and $\Rp = 6$~cm, respectively.}
\label{fig:dens_dis_highff}
\end{figure*}

To understand how fragments might evolve following subsequent collisions we consider what the porosities of the fragments would be following a destructive collision.  Figure~\ref{fig:dens_dis_lowff} (left panel) shows the distribution of fragments with filling factor for the simulations that are just on the fragmenting side of the velocity threshold for initial filling factors $\phi = 15\%$, $25\%$ and $35\%$ (i.e. $\vc = 7.5$, 11 and 12.5 m/s, respectively).  For aggregates with initial filling factor, $\phi = 15\%$, the resulting aggregates are more dense than the initial aggregates since at such low filling factors any interaction leads to compression.  However, for aggregates with filling factors, $\phi = 25\%$ and $35\%$, a bimodal distribution occurs roughly either side of the initial filling factor \citep[as observed by][]{Four_population}, where some aggregates have a smaller filling factor (likely to be due to parts of the aggregate being under tension when they are being pulled off) while other fragments are compressed before being broken off.  Figure~\ref{fig:dens_dis_lowff} (right panel) shows how the mass is distributed amongst the filling factor bins and clearly shows that for the simulations just on the fragmenting side of the threshold velocity, the majority of the mass is at higher filling factors.  However, Figure~\ref{fig:dens_0.35_high_vcoll} shows that as the collision velocity is increased to much higher values than the threshold velocity, a larger proportion of the fragments (and a larger proportion of the mass of the fragments) end up in the lower filling factor bins (likely to be due to the occurrence of more chipping rather than the aggregates being more compressed before breaking off).  In addition, not only does a larger proportion of the mass of the fragments end up in the lower filling factor bin, but the proportion of mass in the higher filling factor bin also reduces as the collision velocity increases.

The fact that the proportion of aggregates in the $> 40\%$ bin is reduced is good for subsequent growth since their fragmentation velocities would be higher than if they had ended up in the higher filling factor bin.

For compact aggregates with initial filling factor $\phi > 37\%$, a decrease in aggregate filling factor does occur and a bimodal distribution is somewhat evident (Figure~\ref{fig:dens_dis_highff}, left panel) though most of the fragments do not increase their filling factor.  This is because these aggregates are initially too brittle and there is little room for compression.  Figure~\ref{fig:dens_dis_highff} (right panel) shows that most of the mass remains in fragments of the same filling factor as that of the original aggregates.  We have verified that this is indeed the case at different collision velocities.

\subsubsection{Porosity of largest aggregate following both growth and destructive collisions}

\begin{figure}
\centering
  \includegraphics[width=1.0\columnwidth]{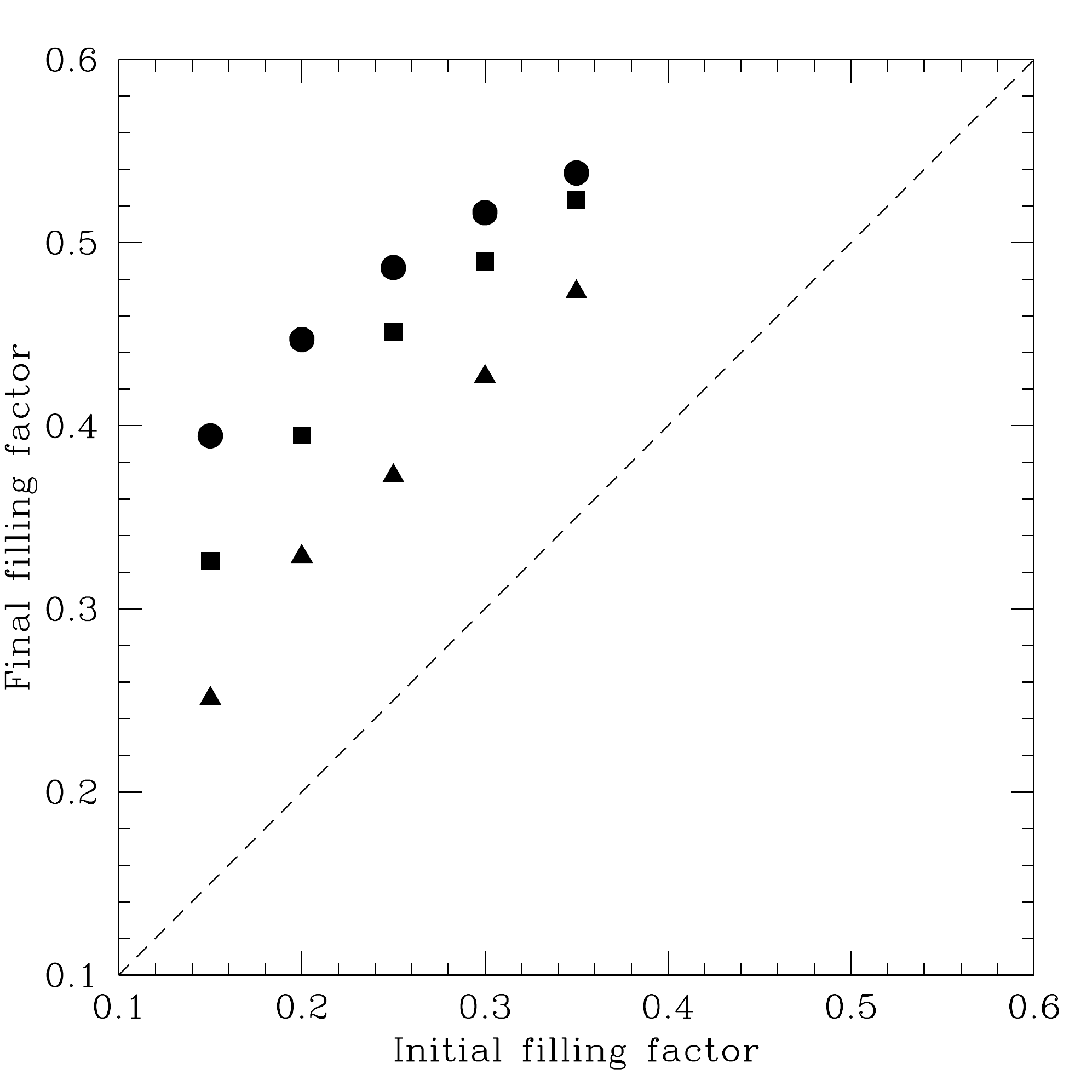}
  \caption{Graph showing the average filling factor of the largest aggregate against initial filling factor for those simulations that are just on the growth side of the fragmentation velocity, with projectile radii 4cm (traingles), 6cm (squares) and 8cm (circles).  The average filling factors increase significantly compared to their initial filling factors.  The initial target radii are $\Rt = 10$~cm.  The dashed line shows the line where the initial and final filling factors are the same.}
\label{fig:ff_mlgst_grow}
\end{figure}

\begin{figure}
\centering
  \includegraphics[width=1.0\columnwidth]{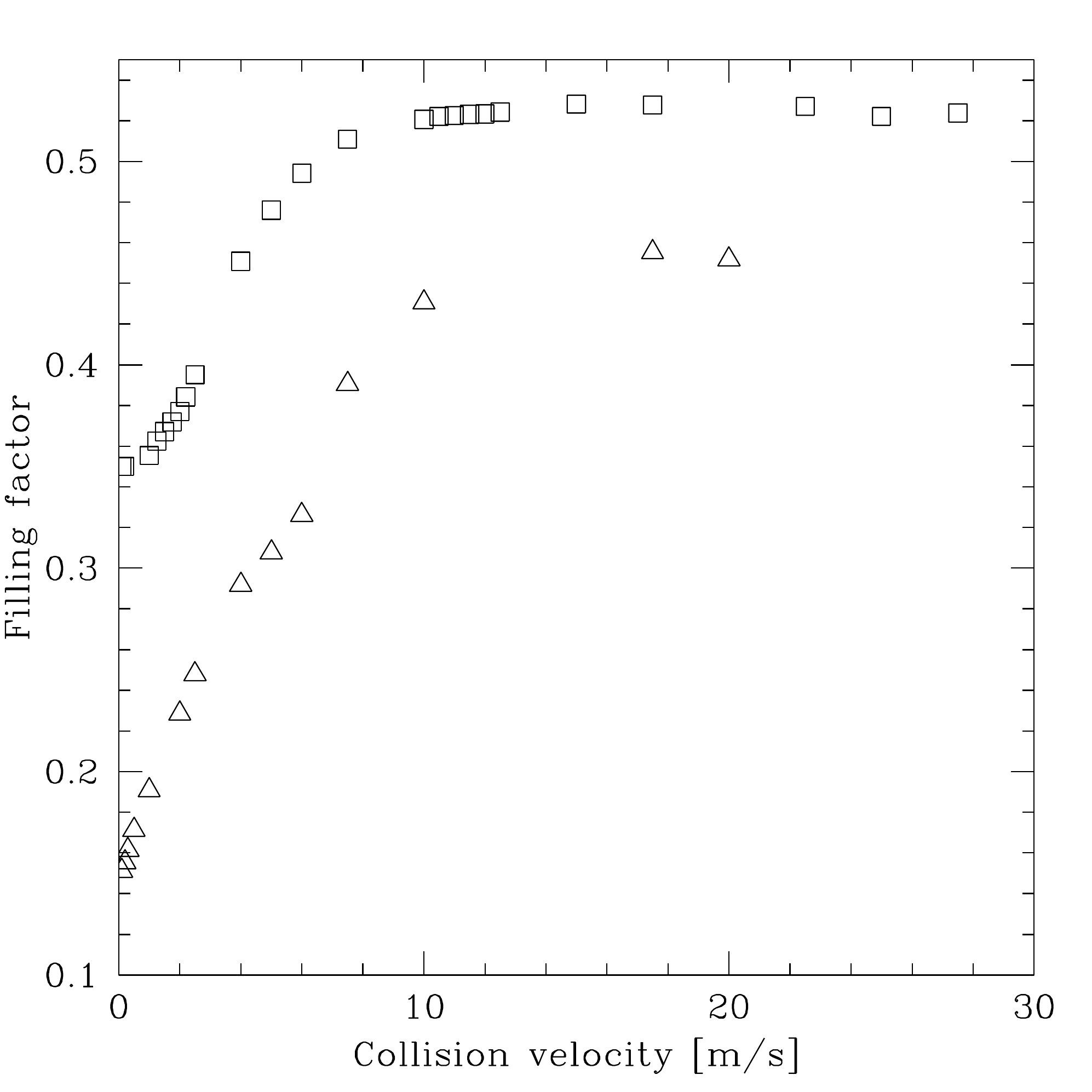}
  \caption{Graph showing the average filling factor of the largest aggregate against collision velocity for simulations carried out with initial filling factors of $\phi = 35\%$ (squares) and 15\% (triangles).  As the collision velocity increases, the final filling factors also increase.  The initial aggregate radii are $\Rt = 10$~cm and $\Rp = 6$~cm.}
\label{fig:ff_mlgst_all}
\end{figure}

During a collision the filling factor of a growing aggregate will increase due to the compaction resulting from the collision.  This will therefore change its ability to withstand future collisions and thus will change its fragmentation velocity.  We calculate the average filling factor of the largest object following both destructive and growth collisions.  This is carried out in a very simplified way (by averaging the density over all the SPH particles) purely to obtain an indication of the final filling factor of the aggregate.  Figure~\ref{fig:ff_mlgst_grow} shows the final filling factors of the largest aggregate in those simulations that are just on the growth side of the fragmentation velocity for initial filling factors $< 37\%$ for projectile radii, $\Rp = 4$, 6 and 8cm.  In all the simulations, a significant increase in filling factor occurs ($> 0.1$) with a larger filling factor increase as the projectile size is increased.  Figure~\ref{fig:ff_mlgst_all} shows how the average filling factor of the largest aggregate varies with increasing collision velocity for aggregates with initial filling factors 15\% and 35\%.  It can be seen that a saturation in filling factor occurs at higher velocities.  This may be expected since during a collision, once fragmentation sets in (at whatever velocity the aggregates are being collided at), no further compression can take place.  Note that since we plot the results for a range of collision velocities, Figure~\ref{fig:ff_mlgst_all} includes the results from both growth and destructive collisions.  At all but the very low velocities, the filling factor of the largest aggregate increases.  These results certainly suggest that subsequent collisions will cause an aggregate to become compacted.  Once it is compacted such that the filling factor of the structure (or perhaps even parts of the structure) have filling factors $\gtrsim 37\%$, we expect the aggregate to have a structure that can be more easily destroyed (i.e. its fragmentation velocity will be lowered due to the existence of high filling factor regions).

We note that only parts of the aggregates will have become compressed - there may be other regions where the filling factors do not change significantly and as a result, collisions with those parts of the aggregate may not lead to destruction.  Figures~\ref{fig:size_illus}~b and c show filling factor rendered images of a growing aggregate.  It can easily be seen that there are regions where the final filling factor is similar to the initial value (see insets) while the region where the impact occurs is much more compacted.  We therefore stress that these results should be taken as indicative.

\section{Discussion}
\label{sec:disc}

\begin{figure}
\centering
  \includegraphics[width=1.0\columnwidth]{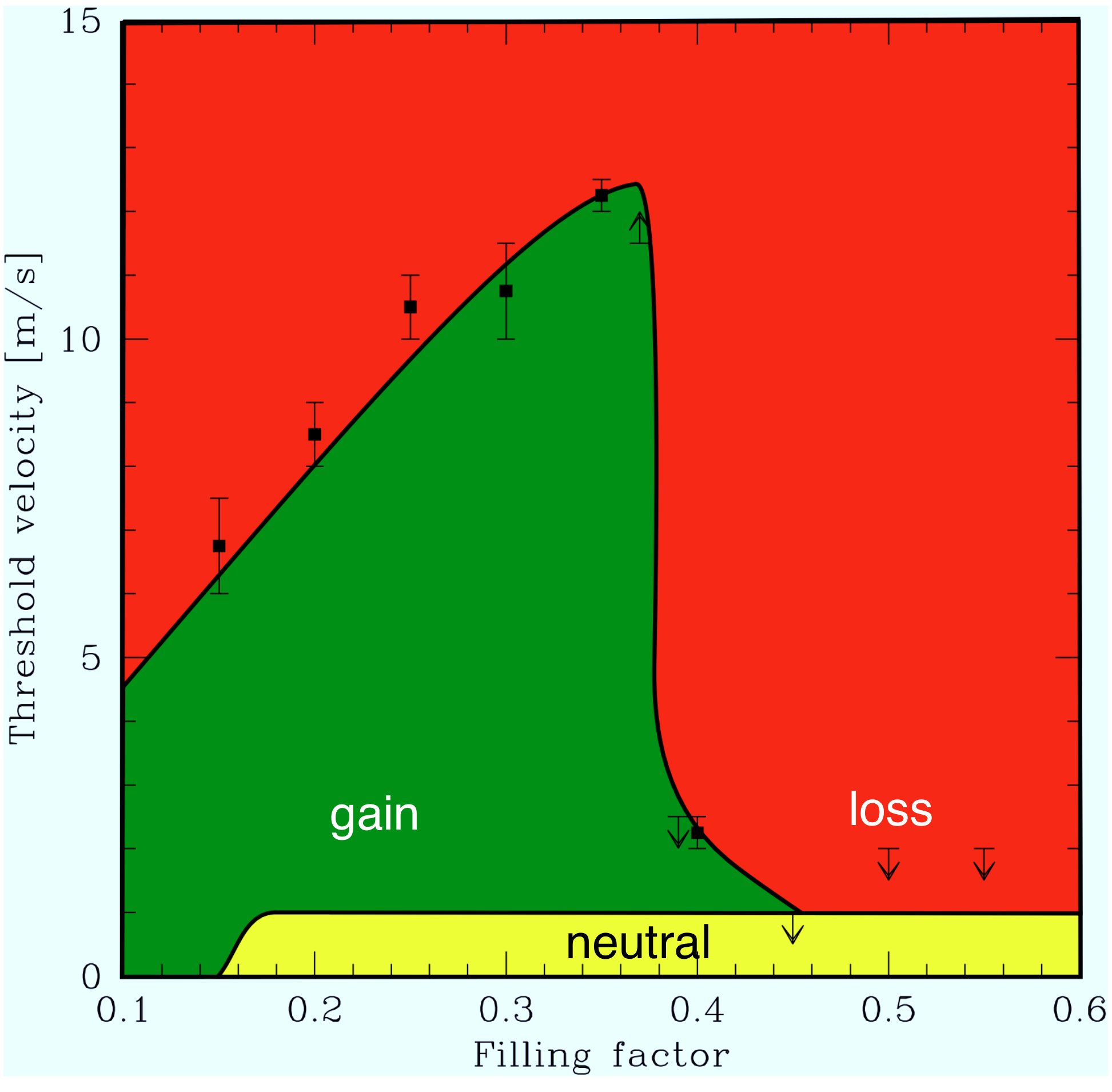}
  \caption{Schematic diagram showing which regions of the collision velocity-filling factor parameter space are expected to result in mass gain, mass loss or a neutral regime upon a collision of two dust aggregates with similar filling factors.  The diagram combines the results presented in this paper (using $\Rt = 10$cm and $\Rp = 6$cm) with those of \citet{Four_population}, \citet{Geretshauser.2012} and \citet{Beitz_bouncing}.}
\label{fig:sketch}
\end{figure}

The growth of dust aggregates in a protoplanetary disc to form planetesimals is a complex process to understand with multiple different aggregate properties that can contribute to the outcome of a collision.  Such information on the effects of these parameters is crucial to understand what conditions are favourable for growth to planetesimal sizes in such discs and in particular how larger dust coagulation/fragmentation models should treat the many parameters that can determine the growth outcome in a protoplanetary disc.  We focus on the effects that the aggregate size and porosity can have on the growth of the aggregate with particular emphasis on when aggregates will be destroyed, and find that the consideration of these properties (and in particular, the effect of aggregate porosity) are very important in order to truly understand the outcome of collisions.  First and foremost, we find that a fixed collision velocity of 1m/s is not the threshold for fragmentation as is often assumed to be the case, but that indeed the outcome does depend on the aggregate properties.  Note that while some authors may refer to a threshold velocity for fragmentation as the velocity at which one of the aggregates may break apart, throughout this paper we refer to it as the smallest velocity at which the largest aggregate remaining after the collision is smaller than the initial target mass.

We carry out an investigation into the effect that the aggregate masses (or equivalently, sizes) have on the fragmentation velocity and find that, for a given density, as the projectile or target sizes are changed (such that the objects' size ratio becomes closer to unity), the threshold velocity decreases.  This suggests that if two objects of different sizes collide, the likelihood of coalescing is higher than for a collision between similar sized objects.

Indeed, in earlier experimental work, \cite{Blum_Munch_1m_s} found that for collisions between equal sized $\rm ZrSiO_4$ aggregates with porosities of $74\%$ and collision velocities $\gtrsim 1$~m/s, fragmentation was the dominant process, whereas it was not the dominant outcome in the equivalent experiments with an aggregate mass ratio of 66, i.e. the collisions with unequal mass objects were able to withstand higher collision velocities (consistent with our results).  In addition, \cite{Wurm_25m/s_impacts} found that using cm-sized targets and mm-sized projectiles made up of $\SiOtwo$ dust with filling factor, $\phi \approx 34\%$, growth occurred even for collision velocities as high as $\approx 25$~m/s (though note that a supporting tray was used for the target and the authors do mention that the results may be different if such a tray was not used).  More recently, \cite{Teiser_Wurm_highVcoll} find that growth can occur even when collision velocities are as high as $\approx 55$~m/s for a projectile that is as small as $\approx 0.5$~mm-sized against a target that is at least 10 times larger and has a filling factor, $\phi \approx 33\%$.  Furthermore, we show using numerical simulations that for two aggregates that are both in the centimetre to decimetre size regime, growth may in some cases occur even when the collision velocities are at least as high as $\approx 27.5$ m/s.  We also determine the dependence of the threshold velocity for fragmentation on the projectile mass and radius, and show that this dependence is of the form $v_{\rm th} \propto \Rp^{\sigma_{\rm r}}$ (or equivalently, $v_{\rm th} \propto \mproj^{\sigma_{\rm m}}$), for the explored filling factor range of up to 35\%.  Such trends can be used in coagulation and fragmentation simulations to accurately understand the growth of particles.

In addition, we find that even if the mass ratio of the aggregates is kept constant, the smaller sized objects are able to withstand higher velocities and therefore it is far easier for small aggregates to grow.  However, the difficulty arises when they grow larger: the aggregate velocities in a protoplanetary disc will increase yet the threshold velocity for fragmentation for two objects with the same mass ratio will decrease.  Therefore the only avenue for growth for an object at any one location in a disc is for collisions to occur with aggregates much smaller in size 
(\citealp{Teiser_Wurm_highVcoll,Teiser_Wurm_decimeter_growth}; \citealp{Windmark_lucky_ptcl}).  This also suggests that even if a destructive collision was to occur, as long as some larger objects remain, growth may still be enhanced in a following collision if the size ratio of the objects is large.  It is well known that in a destructive collision, fragments form such that the cumulative distribution of their fragment masses is a power-law of the form, $m_{\rm cum} \propto m_{\rm f}^\kappa$, where $m_{\rm cum}$ and $m_{\rm f}$ are the cumulative and fragment masses, respectively.  $\kappa$ is the power-law index which \cite{Four_population} show to decrease as the collision velocity increases, i.e. in a more destructive collision many more low mass fragments form.  Regardless of the magnitude of $\kappa$, a size distribution always results and thus the small objects may then collide with the larger objects, which may lead to growth.  Thus, while fragmentation at first seems to be a destructive process, it may end up aiding the growth of aggregates.  We note that our results on collisions between aggregates with a constant mass ratio is in the opposite sense to \cite{Wada_icy_collisions} who find that the larger aggregates are able to withstand higher collision velocities.  However, it must be noted that the regimes are quite different: our aggregates are much larger in size where the effects of plastic deformation are more important \citep{Chokshi_vs} while the simulations by \cite{Wada_icy_collisions} were of much smaller aggregates and the effect of plastic deformation was not taken into account.

Crucially, we find that the porosity of the aggregates plays an
extremely important role in whether they can not only survive a
collision, but also grow.  We find a regime of parameter space above a
filling factor of $\approx 37\%$ in which the growth regime is
significantly reduced and perhaps even non-existent.  We find that
above $\approx 2$~m/s such compact aggregates break apart while they
exhibit neutral behaviour (i.e. they rebound off each other) when the
collision velocities are at or below this value.  This is consistent
with the simulation results presented by \cite{Four_population} and
\cite{Geretshauser.2012} as well as the results obtained by
\cite{Beitz_bouncing} who carry out laboratory collisions of
centimetre-sized $\SiOtwo$ dust aggregates with filling factors $>
35\%$ and find that fragmentation occurs for collision velocities
higher than $1.9$~m/s.  Below this value, they find either bouncing or
what they describe as ``partial bouncing with mass transfer''.  We
note that the latter describes simulations where the accretion
efficiency (increase in mass in units of the projectile mass) is
approximately a few percent, which we class as being in the neutral
regime (equations~\ref{eq:neutral_ml} and~\ref{eq:neutral_ms}).
Furthermore, our results are also consistent with the high
filling factor results of \cite{Schraepler_5cm_compact} who carry out
collisions between 5cm-sized dust aggregates.  They find a velocity
threshold of $\approx 1$m/s below which bouncing occurs (see their
Figure 6).  Their results for $\phi = 30\%$ also suggest a velocity
threshold of $\approx 1$m/s (note that they do not have a data point
that strictly speaking confirms this), which at first glance appears
to be inconsistent with our results.  However, the authors note that
the preparation of their sample involved inhomogeneous compression
such that the filling factors in some parts may have been as high as
$40\%$.  Our results show that an aggregate with filling factor
$\gtrsim 37\%$ is much easier to destroy.  Therefore, the introduction
of compressed regions (with $\phi > 37\%$) may therefore have
introduced weaker regions into the sample, resulting in a lower
threshold velocity in the experiments.  Note that the present value of
the filling factor $\phi = 37\%$, where the fragmentation velocity drops
significantly, depends on the material properties. However, we
stress that such a drop always has to occur when the porosity limit
is reached by compaction, resulting in the compressive strength increasing
indefinitely.  Therefore, the actual limit (i.e. the drop in
fragmentation velocity) may be at different filling factors for
different materials such as polydisperse $\SiOtwo$.

Below the filling factor value of $\approx 37\%$, a growth region exists and the size of this region increases with the filling factor.  \cite{Wurm_mm_cm} carried out laboratory experiments (using $\SiOtwo$) of mm-sized aggregates colliding with cm-sized targets with filling factors between 12\% and 26\% and found that material was ejected during the impact, while \cite{Wurm_25m/s_impacts} carried out similar experiments using aggregates with filling factors of $\approx 34\%$ with collision velocities of up to 25m/s and found that these were able to grow.  These results already suggest that a higher filling factor can withstand a higher collision velocity, a result that we confirm for aggregates with filling factors $\lesssim 37\%$.  Furthermore, our results are in qualitative agreement with \cite{Wada_icy_collisions} who carry out numerical simulations of collisions between icy clusters.  They find that more porous aggregates (formed via balistic particle-cluster aggregation) are more easily destroyed than less porous ones (formed via balistic cluster-cluster aggregation).

Experimental results by \cite{Love_high_vel_expts} and \cite{Jutzi_high_v_SPH} of collisions at higher velocities ($\approx \rm O(1)$~km/s) indicated that lower filling factor objects may withstand higher collision velocities.  However, it is important to note that the processes involved in such high velocity collisions are different to those at lower velocity such as those simulated in this paper (e.g. the effects of melting become important at higher velocities and thus higher energies).  We also note that the expected collision velocities for centimetre- to decimetre-sized objects in protoplanetary discs ranges between 1 m/s to $\approx 100$~m/s \citep{Weidenschilling1977}, i.e. much lower than those considered by \cite{Love_high_vel_expts} and \cite{Jutzi_high_v_SPH}.  Furthermore, \cite{Love_high_vel_expts} used lime glass beads which were sintered and thus have different strength properties.  In addition, their experiments involved aggregates with $\phi > 40\%$ and so their conclusions on the collisional outcome are for a different regime to ours ($\phi \lesssim 37\%$).

Following a collision, we find that the largest aggregate is on average compressed during the collision (in agreement with laboratory experiments by \citealp{Teiser.2011} and \citealp{Kothe_ff_vcoll_relation}).  While this might help to stabilise the aggregate to a certain extent, if regions of the aggregate become compressed such that their filling factors increase to $\gtrsim 37\%$, the aggregate will be weakened.  However, we also find that in a destructive collision the resulting fragments are a mixture of both porous and compact aggregates.  As the collision velocity is increased to much higher values than the fragmentation velocities, the proportion of low filling factor (i.e. $\phi \lesssim 37\%$) aggregates increases.  These may then go on to play a ``growth role'' in a future collision.  However, collisions between different porosity aggregates need to be simulated to fully understand the collisional outcomes.

Combining the key porosity results for centimetre-sized objects presented in this paper with those of \cite{Four_population}, \cite{Geretshauser.2012} and \cite{Beitz_bouncing}, we expect the gain, loss and neutral regimes to be approximately of the form presented in the sketch in Figure~\ref{fig:sketch}.  From such an expectation, it is clear that the highly compact regime presents a problem for the growth of aggregates when roughly equal-sized objects collide.  However from Section~\ref{sec:mass}, since we find that collisions between unequal-sized aggregates are more likely to result in growth, this area of the parameter space may present a possibility for the growth of such compact aggregates.  Coagulation codes that do not consider the detailed effects of porosity may therefore under- or over-estimate the growth of aggregates depending on the fragmentation velocity adopted.  We note that in the sketch in Figure~\ref{fig:sketch} we have not included any growth effects at very low velocities.  However, we would expect this to occur due to the gentle ``hit and stick'' collisions.

\cite{Geretshauser_inhomogeneity} carried out SPH simulations of inhomogeneous $\SiOtwo$ dust aggregates and found that as the degree of inhomogeneity increased the aggregates were more likely to be destroyed.  Figure 2 (upper panel) and 3 of \cite{Geretshauser_inhomogeneity} show that when a significant portion of the aggregates consist of regions with filling factors $\gtrsim 37\%$, the aggregates are broken apart when otherwise they stick together and grow.  As with the experimental results by \cite{Schraepler_5cm_compact}, we expect that the level of destruction of the aggregates in the simulations presented by \cite{Geretshauser_inhomogeneity} is largely due to the introduction of ultra-weak regions in which the filling factor is $\gtrsim 37\%$.  Therefore, while we anticipate that our fragmentation velocities are upper limits since dust aggregates in protoplanetary discs will no doubt be inhomogeneous, we expect that the degree to which the threshold velocity is reduced due to inhomogeneity depends on whether parts of the aggregates have become compressed to filling factors $\gtrsim 37\%$.

\cite{Ringl_vth} carry out molecular dynamics simulations of equal-sized aggregates with radii, $R = 28 \mu \rm m$, and filling factor, $\phi = 20.5\%$.  They show that for head-on collisions, the critical velocity for fragmentation is $\approx 17$~m/s.  While the size range is very different to that which we are considering, it is interesting to note that they find a high threshold velocity for fragmentation (larger than the typically assumed threshold value of $\approx 1$~m/s).

\cite{Windmark_lucky_ptcl} carry out dust coagulation simulations assuming that the aggregates are always compact.  Our results show that such an assumption can neglect an entire growth regime, thus under-estimating growth.  While we do also show that in a fragmenting collision, most of the mass tends to result in more compact fragments (and specifically there is the indication that the largest aggregate becomes more compact during a collision), some objects with lower filling factors still remain, which may end up sticking to other aggregates in subsequent collisions.  On the other hand, \cite{Teiser_Wurm_decimeter_growth} carry out laboratory experiments where decimetre sized objects are grown in the laboratory by colliding projectiles with sizes up to $\approx 250 \mu$m with targets of sizes up to several centimetres.  They find that regardless of the collisional history (which includes both direct impacts at velocities of up to 9.2m/s as well as secondary slower collisions with velocities up to 1m/s due to reaccretion of the ejecta back onto the aggregate), the aggregates end up with filling factors of $\approx 30\%$ which we show is in a regime of the parameter space where growth can occur.  We note that firstly, our porosity investigation involves aggregates that are approximately similar sized while those experiments by \cite{Teiser_Wurm_decimeter_growth} involve very unequal-sized aggregates.  This therefore may be a channel by which growth of aggregates may occur even within the highly compact regime.  Secondly, we note that the porosity of the initial aggregates in our simulations are identical whereas such a controlled test is not possible in the laboratory.  Given that we show the importance of the aggregate porosity, the threshold velocities may well be affected (or perhaps dominated by) the porosity of one of the aggregates.  Nevertheless, it is clear that the effects of the aggregate porosity is an important and complex property.  To fully understand the importance of these lower filling factor aggregates, the effects of aggregate porosity needs to be considered in dust coagulation codes.

Based on a set of 19 experiments, \cite{Guttler.2010} attempted to map out the regions of the projectile mass and collision velocity parameter space where bouncing, sticking and fragmentation occur.  In particular, our simulations can be compared to the first three plots in their Figure 11 (the first row of plots and the first plot in their second row).  Our simulations of collisions between different sized aggregates can be compared to their first row of plots while those carried out with porosities $\phi < 0.4$ and $\phi \ge 0.4$ can be compared to their first and third plots, respectively.  Our results are qualitatively consistent with their expectation that as the projectile mass increases, the velocity above which sticking no longer occurs decreases.  In addition, our results for similar sized compact aggregates are also consistent with theirs in that we do not see a growth region between the neutral and loss regimes.  However, for similar sized porous aggregates they also suggest a transition directly from a neutral regime to a loss regime whereas we clearly see a transition from growth to loss.  Thus, our results open up an area in which growth can occur that was not present in their collision model.

\subsection{Observations and their role in directing future modelling}

Much observational focus has been directed towards understanding the grain sizes and their distribution in protoplanetary discs.  It is clear that sizes ranging from sub-micron to centimetre sizes can exist in such discs (see Section 6.3 of \citealp{Williams_Cieza_disc_review}).  If such variations in size occur at the same location in a protoplanetary disc then our results suggest that growth is more promising.  A recent coagulation/fragmentation study by \cite{Garaud_vel_pdf} suggests the co-existence of two particle populations (also see \citealp{Windmark_lucky_ptcl}).  The net outcome of this in the context of our results is that both small and large sizes are present at any one location in a disc which our results show is favourable for growth.

Pinte et al (2008) carry out observations of the circumstellar disc around IM Lupi to determine the dust grain evolution.  They find that the anisotropic scattering  seen in the scattered light images is at least consistent with the presence of porous grains ($\approx 80\%$).  However, the data is also consistent with the presence of ice mantles around the dust grains, a degeneracy that they state may be solved if the disc is observed in polarised light.  \cite{Perrin_60percent_porous} observe AB Aurigae's disc in polarised light and find that their data is consistent with grains that are $\approx 60\%$ porous, i.e. just on the edge of the region where we find that growth becomes much harder.  Future observations that address the question of whether porous ($\lesssim 37\%$) or compact ($\gtrsim 37\%$) grains exist in protoplanetary discs will help to determine whether the dust in real discs have porosities that allow them to lie in the region of the parameter space where growth is more likely, or in regions which hinder planet formation.  If porous grains do indeed exist in protoplanetary discs, it at least suggests that growth is not severely hindered by the presence of highly compact aggregates.  On the other hand if the observational data is more often only consistent with compact aggregates, the challenge for modellers will be to understand the growth processes despite the fact that growth is severely hindered in this regime.  Such observations will enable future modelling of dust collisions to focus on the porosities most relevant to dust in protoplanetary discs.

\section{Conclusions}
\label{sec:conc}

We carry out high resolution three-dimensional Smoothed Particle Hydrodynamics simulations of collisions between $\SiOtwo$ dust aggregates with radii between 2-15 cm.  We show that the sizes of both the target and projectile play a part in whether the aggregates survive a collision.  We find that aggregates can survive collisions with velocities that are mostly higher than the assumed threshold value of 1m/s and find that objects of very different sizes can survive higher collision velocities -  even as high as at least 27.5m/s.  Therefore, while one must of course bear in mind that the simulations here are ideal cases of head-on spherical homogeneous aggregates, our results suggest the possibility that the parameter space for growth may be larger than previously considered.

Crucially, we also find that the porosity of an aggregate is key to its survivability.  As the filling factor increases, the aggregates can withstand a higher collision velocity and can still grow as their ability to dissipate energy increases, potentially opening up a region of parameter space where growth can occur.  However, we show that growth is significantly hindered at filling factors larger than $\approx 37\%$ as energy can no longer be dissipated efficiently by plastic compression.  Therefore, if aggregates become too compact growth may be hindered.  On the other hand we also find that in fragmenting collisions, some fragments with a lower filling factor do result which may then survive higher collision velocities than compact aggregates in subsequent impacts.

Thus we conclude that a \emph{single fragmentation velocity is insufficient when trying to accurately capture the growth of dust aggregates} and that the properties of both the aggregates are important to consider.

\section*{Acknowledgments}
We would like to thank the anonymous referee, Satoshi Okuzumi, Fredrik Windmark, Hidekazu Tanaka, Koji Wada, Marina Galvagni and Jean-Fran\c{c}ois Gonzalez for useful comments.  The calculations reported here were performed using the university
cluster of the computing centre of T\"ubingen and the bwGRiD clusters in
Karlsruhe, Stuttgart, and T\"ubingen. Computing time was also provided
by the High Performance Computing Centre Stuttgart (HLRS) on the
national supercomputer NEC Nehalem Cluster under project grant
SPH-PPC/12848.  We acknowledge the support of the German Research Foundation (DFG) through grant KL 650/8-2 within the Collaborative Research Group FOR 759: {\it The formation of Planets: The Critical First Growth Phase}.  FM is supported by the ETH Zurich Postdoctoral Fellowship Program as well as by the Marie Curie Actions for People COFUND program.

\bibliographystyle{mn2e}
\bibliography{allpapers}

\appendix
\section{Numerical tests}
\label{appendix}

We re-run the Reference simulation but change the initial setup so that we use a hexagonal lattice.  These simulations are run using collision velocities of 12.0 and 12.5 m/s.  We find that the former collision velocity allows growth while the latter results in fragmentation.  Furthermore, we also simulate the Reference case but change the alignment of the SPH particles such that: (i) the SPH particles in the target are initially aligned with the collision axis while those in the projectile are initially rotated by $20^{\circ}$, and (ii) the SPH particles in the target and projectile are initially rotated by $10^{\circ}$ and $20^{\circ}$, respectively.  These simulations are run using collision velocities of 12.5 m/s which result in growth, and 13.5 m/s which result in fragmentation.  These results for the velocity thresholds are similar to that obtained for the Reference case ($12.25 \pm 0.25$; see Section~\ref{sec:results}).

\end{document}